\documentclass{emulateapj-rtx4}
\usepackage{amssymb}
\usepackage{natbib}
\usepackage{lscape}
\bibstyle{aa}
\shorttitle{Diffuse Lyman Alpha Halos}
\shortauthors{Steidel et al.}

\newcommand{\lsun}{\ensuremath{\rm L_\odot}}

\newcommand{\Ha}{\ensuremath{\rm H\alpha}}
\newcommand{\Hb}{\ensuremath{\rm H\beta}}
\newcommand{\lya}{\ensuremath{\rm Ly\alpha}}

\newcommand{\lyb}{Ly$\beta$}

\newcommand{\minpoint}{\mbox{$'\mskip-4.7mu.\mskip0.8mu$}}

\newcommand{\secpoint}{\mbox{$''\mskip-7.6mu.\,$}}

\def\ltsima{$\; \buildrel < \over \sim \;$}
\def\simlt{\lower.5ex\hbox{\ltsima}}
\def\gtsima{$\; \buildrel > \over \sim \;$}
\def\simgt{\lower.5ex\hbox{\gtsima}}
\def\arcs{$''~$}

\begin{document}
\title{DIFFUSE LYMAN ALPHA EMITTING HALOS: A GENERIC PROPERTY OF HIGH REDSHIFT STAR FORMING GALAXIES\altaffilmark{1}}
\slugcomment{Accepted for publication in the Astrophysical Journal}
\author{\sc Charles C. Steidel\altaffilmark{2},  Milan Bogosavljevi\'c\altaffilmark{2,3}, Alice E. Shapley \altaffilmark{4,5,6}, \\
Juna A. Kollmeier\altaffilmark{7}, Naveen A. Reddy\altaffilmark{8,9}, Dawn K. Erb\altaffilmark{10,11,12}, \& Max Pettini\altaffilmark{13}  } 

\altaffiltext{1}{Based on data obtained at the 
W.M. Keck Observatory, which 
is operated as a scientific partnership among the California Institute of 
Technology, the
University of California, and NASA, and was made possible by the generous 
financial
support of the W.M. Keck Foundation. 
}
\altaffiltext{2}{California Institute of Technology, MS 249-17, Pasadena, CA 91125, USA}
\altaffiltext{3}{Astronomical Observatory, Volgina 7, 11060 Belgrade, Serbia}
\altaffiltext{4}{Department of Physics and Astronomy, University of California, Los Angeles, 430 Portola
Plaza, Box 951547, Los Angeles, CA 90095, USA}
\altaffiltext{5}{Alfred P. Sloan Fellow}
\altaffiltext{6}{Packard Fellow}
\altaffiltext{7}{Carnegie Observatories, 813 Santa Barbara Street, Pasadena, CA 91101, USA}
\altaffiltext{8}{National Optical Astronomy Observatories, 950 N. Cherry Ave., Tucson, AZ 85258, USA}
\altaffiltext{9}{Hubble Fellow}
\altaffiltext{10}{Department of Physics, University of California, Santa Barbara, Santa Barbara,
CA 93106, USA}
\altaffiltext{11}{Spitzer Fellow}
\altaffiltext{12}{Department of Physics, University of Wisconsin-Milwaukee, P.O. Box 413, Milwaukee, WI 53201, USA}
\altaffiltext{13}{Institute of Astronomy, Madingley Road, Cambridge CB3 OHA UK}
\setcounter{footnote}{0}
\begin{abstract}
Using a sample of 92 UV continuum-selected, spectroscopically identified galaxies with $\langle z \rangle = 2.65$,
all of which have been imaged in the \lya\ line with extremely deep narrow-band imaging, 
we examine galaxy \lya\ emission profiles to very faint surface brightness limits. 
The galaxy sample is representative of spectroscopic samples
of LBGs at similar redshifts in terms of apparent magnitude, UV luminosity, inferred extinction, 
and star formation rate and was assembled without regard to \lya\ emission properties.
Approximately 45\% (55\%) of the galaxy spectra have \lya\ appearing in net absorption (emission),
with $\simeq 20$\% satisfying commonly used criteria for the identification of ``Lyman Alpha Emitters'' (LAEs) [$W_0(\lya) \ge 20$ \AA].
We use extremely deep stacks of rest-UV continuum 
and continuum-subtracted \lya\ images to show that all sub-samples exhibit diffuse 
\lya\ emission to radii of at least 10\arcs\ ($\sim 80$ physical kpc). The characteristic exponential 
scale lengths for \lya\ line emission exceed that of the $\lambda_0=1220$ \AA\ UV continuum light by 
factors of $\sim 5-10$.
The surface brightness profiles of \lya\ emission are strongly suppressed relative to the UV continuum light in
the inner few kpc, by amounts that are tightly correlated with the galaxies' observed spectral morphology; 
however, all galaxy sub-subsamples, including that of galaxies for which \lya\ appears in net absorption in 
the spectra, exhibit qualitatively similar diffuse \lya\ emission halos.  
Accounting for the extended \lya\ emission halos, which generally would not be detected in the slit spectra of
individual objects or with typical narrow-band \lya\ imaging, 
increases the total \lya\ flux [and rest equivalent width $W_0(\lya)$] 
by an average factor of $\sim 5$, and by a much larger factor for the 80\% of LBGs not classified as LAEs. 
We argue that most, if not all, of the observed \lya\ emission in the diffuse halos originates in the
galaxy \ion{H}{2} regions but is scattered in our direction by \ion{H}{1} gas in the galaxy's circum-galactic medium (CGM). 
The overall intensity of \lya\ halos, but not the surface brightness distribution,  is strongly correlated
with the emission observed in the central $\sim 1$\arcs\-- more luminous halos are observed for galaxies with
stronger central \lya\ emission.  We show that whether or not a galaxy is
classified as a giant ``Lyman $\alpha$ Blob'' (LAB) depends sensitively on the \lya\ surface brightness threshold
reached by an observation. Accounting for diffuse \lya\ halos, all LBGs would be LABs if 
surveys were sensitive to 10 times lower \lya\ surface brightness thresholds; similarly, 
essentially all LBGs would qualify as LAEs.  
\end{abstract}

\section{Introduction}

Although the Lyman $\alpha$ (\lya) emission line of neutral H is expected 
to be produced in prodigious amounts by star-forming galaxies (e.g. \citealt{partridge67,meier76a}), 
it has long been appreciated
that the astrophysics affecting observations of \lya\ are far more complex than for other lines of 
abundant species due to resonant scattering (\citealt{spitzer78,meier81,charlot93}). 
The very large cross-section in the \lya\ 
transition means that emission from a gas cloud or nebula 
may have been strongly altered in intensity, kinematics, and apparent spatial distribution by
the time it reaches an observer. Similarly, information about the 
initial source of observed \lya\ emission may be lost or obscured, with the apparent
source simply being \ion{H}{1} gas responsible for scattering in the observer's direction. 
Consequently, the dominant process producing \lya\ emission may often be 
ambiguous; possibilities include 
photoionization by young stars or AGN, line emission following collisional
excitation of H atoms, 
or simply scattering from intervening \ion{H}{1} gas that
happens to favor the observer's direction. 

In the absence of dust, the standard expectation for \lya\ emission produced 
in \ion{H}{2} regions for ``Case B'' (i.e., ionization-bounded) 
recombination \citep{brockle71} and a \cite{chabrier03} stellar initial mass function (IMF)
for high mass stars
\footnote{Note that this value is a factor 1.8 higher
than would be obtained assuming a Salpeter (1955) IMF because a given number of ionizing photons
is associated with a smaller total SFR for the Chabrier IMF.}
is that each solar mass of star formation produces a \lya\ luminosity
$L(\lya) \simeq 2.0 \times 10^{42}$ ergs s$^{-1}$.  
For the same IMF, the far-UV continuum light produced per solar mass of SFR near the wavelength of \lya\ 
has an expected  monochromatic luminosity in the range $40.0 \simlt{\rm log}~L_{\lambda,cont} \simlt 40.3$ ergs s$^{-1}$
\AA$^{-1}$  (\citealt{leitherer99}). 
The predicted rest equivalent width of \lya\ emission is then given by
$W_0(\lya) \simeq L(\lya)/L_{\lambda,cont} \simeq 100 - 200$ \AA\ (see also \citealt{charlot93}),
with values near the lower end of this range expected for continuous star formation lasting 
more than $\sim 3 \times 10^7$ yrs, roughly the minimum dynamical timescale for L* LBGs at $z \sim 2-3$ (e.g.,
\citealt{erb+06b}). Under the above assumptions, the period of time over which 
\lya\ emission has $W_0(\lya) > 100$ \AA\ would be very brief, after which the line-to-continuum ratio
reaches an asymptotic value of  
$W_0(\lya) \simeq 100$ \AA. Thus, for a UV continuum-selected sample, one would expect only a small fraction of  
galaxies to be caught during a time when their intrinsic $W_0(\lya)$ exceeds 100 \AA \footnote{For a sample 
selected by \lya\  (as opposed to continuum) emission, this may not be the case.}. 

When dust is mixed throughout the scattering medium,
one expects selective extinction of \lya\ photons compared to those in the nearby UV continuum
due to the much larger effective path length traversed by a line photon before 
escaping into the intergalactic medium
(e.g., \citealt{meier81,hartmann84,neufeld90}). 
This effect is often cited when observed \lya\ emission lines are much weaker 
than the Case B expectations discussed above (e.g., \citealt{charlot93,shapley03,hayes10,kornei10}). 
Since most continuum-selected high redshift galaxies in current spectroscopic surveys 
appear to have at least some dust, and the vast majority have \lya\ equivalent widths 
$W_0(\lya) < 100$ \AA\ (e.g., \citealt{shapley03,kornei10}), this conclusion would seem reasonable. 
On the other hand, it is also possible, at least in principle, for \lya\ photons to experience {\it less} attenuation  
by dust than continuum photons, in the case of a clumpy ISM in which dust is located only
within the clumps which are rarely penetrated by \lya\ photons (\citealt{neufeld91, finkelstein08}). 
There is no reason to believe that the two competing effects 
could not
{\it both} be at work within different regions of the same galaxy.  

Even without dust, however, resonant scattering produces spatial and/or spectral diffusion 
of \lya\ photons leading to emergent line emission whose properties depend on the geometry, kinematics, 
and \ion{H}{1} optical
depth distributions within the gaseous circumgalactic medium (CGM) surrounding a galaxy 
(\citealt{steidel2010} [S2010]). In the zero-dust case, the total \lya\ luminosity would be unaltered 
by resonant scattering, but, as we detail below, 
the {\it detectability} of \lya\ could be very strongly affected.

\begin{figure}[thb]
\centerline{\includegraphics[height=9cm]{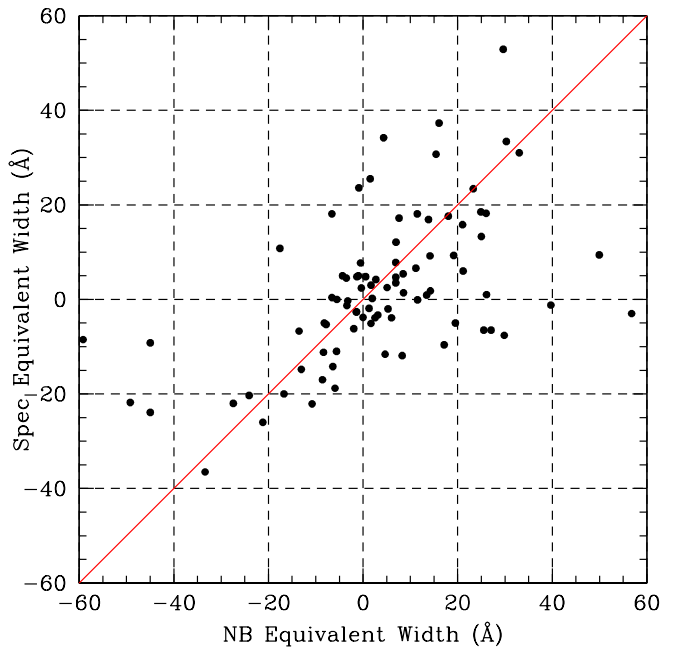}}
\caption{ Comparison of \lya\ line equivalent widths measured from spectra compared to those
inferred from Cont-NB colors in deep\lya\ imaging. The imaging measurements use
isophotal apertures defined by
the extent of \lya\ flux to a surface brightness limit of $\simeq 1-2 \times 10^{-18}$
ergs s$^{-1}$ cm$^{-2}$ arcsec$^{-2}$, which is typical of the deepest \lya\ narrow-band imaging surveys.
\label{fig:ew_vs_ew}
}
\end{figure}

\begin{figure}[thb]
\centerline{\includegraphics[height=9cm]{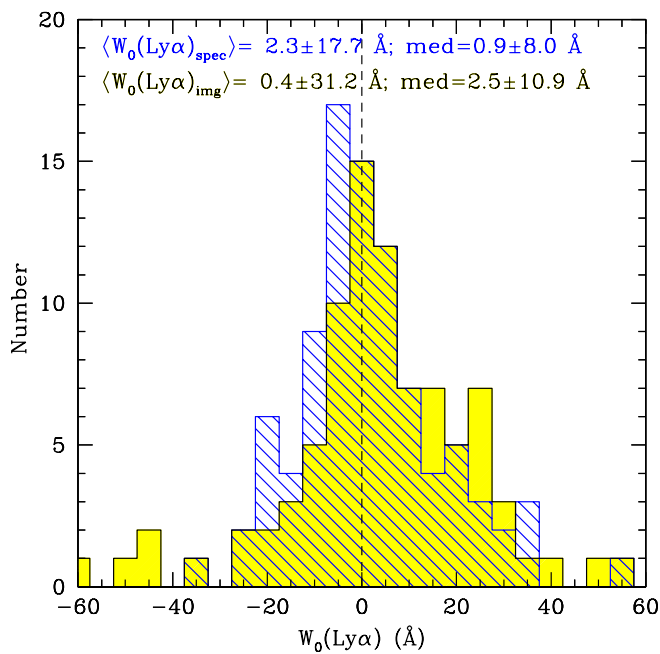}}
\caption{Comparison of the \lya\ line equivalent width distribution from spectroscopic measurements
versus that inferred from CB-NB colors in \lya\ imaging. The imaging measurements use colors
within isophotal apertures defined by     
the extent of \lya\ flux to a surface brightness limit of $\simeq 1-2 \times 10^{-18}$
ergs s$^{-1}$ cm$^{-2}$ arcsec$^{-2}$, which is typical of the deepest \lya\ narrow-band imaging surveys.  
The statistics are for the mean and standard deviation (left) of individual values (left), and the
median and inter-quartile range (right) for each set of measurements.  
\label{fig:ew_hist}
}
\end{figure}

In S2010, we 
characterized the distribution of cool gas in the CGM of star-forming galaxies with redshifts 
$2 \simlt z \simlt 3$ and attempted to
understand the kinematics and line strength of the ISM absorption and \lya\ emission in the context
of galaxy-scale gaseous outflows. 
In brief, we found that UV-selected galaxies within a factor of a few of L* in the far-UV continuum luminosity function
(corresponding at $z \sim 2.5$ to apparent magnitudes ${\cal R} \simeq 24-24.5$-- see \citealt{reddy09})
have a CGM that can be traced by \ion{H}{1} (\lya\ and \lyb\ absorption) and several strong
absorption lines of metallic species (e.g., \ion{C}{2}, \ion{C}{4}, \ion{Si}{2}, \ion{Si}{4}) to galactocentric
distances of $\simlt 120$ kpc using the spectra of faint background galaxies.  The measurement used more than 
500 galaxy pairs on angular scales $1-15$\arcs\ to map out the absorption line strength as a function of
galaxy impact parameter $b$ (i.e., the physical separation of the two lines of sight
at the redshift of the foreground galaxy) for each observed species. 
In slit spectra of the CGM ``host galaxies'', the bulk of observed \lya\ emission, when present,
is almost always strongly redshifted, while the strong interstellar (IS) absorption lines are strongly blue-shifted. 
S2010 presented a geometric and kinematic model that reproduces many of the observed trends. 

In the context of the model, \lya\ photons escape the galaxy 
in an observer's direction mainly 
by scattering from optically thick \ion{H}{1} gas located on the far
side of the galaxy's stars, but having the same overall (outflowing) kinematics as the 
IS gas seen in blue-shifted absorption. 
We used the transverse information
from the galaxy pairs combined with line-of-sight information available from the galaxies' own
far-UV spectra to construct a consistent geometric and kinematic model of galaxy scale outflows in the context of
a very well-studied population of high redshift star-forming galaxies. That is, 
we combined the line profiles of IS absorption lines and \lya\ emission in the galaxy spectra themselves (sampling the
kinematics and line strength for 
galactocentric impact parameter $b \sim 0$) with IS line strength measurements at $b>>0$ (using close
angular pairs of galaxies) to infer the 3-dimensional
distribution of CGM gas surrounding an average galaxy in the spectroscopic sample.  
We suggested that the CGM gas seen in absorption would also constitute a scattering medium through
which \lya\ photons must traverse in order to be observed. High velocities and large velocity gradients 
together with gas covering fraction $f_c \le 1$ through much of the CGM allow \lya\ photons
to diffuse spatially outward, favoring escape of \lya\ photons last scattered (in the observer's direction) 
from atoms with velocities well off resonance with respect to any \ion{H}{1} that remains between the
location of the last scattering and the observer.  If true, one might then expect to observe  
scattered \lya\ emission over the same spatial scales for which strong HI and low-ion metallic {\it absorption}
is seen, i.e., $\simeq 80-90$ kpc, even if all \lya\ photons originated in the
galaxy's \ion{H}{2} regions. 

Clearly, scattering will substantially modify both the spatial and spectral distribution of \lya\ photons
emergent in a particular direction, and at the very least may cause \lya\ emitting regions to appear
distinct from the UV continuum emission even if both share a common origin. Slit spectra commonly optimized
for the compact size of the continuum emitting regions of typical star-forming high redshift galaxies 
may encompass only a fraction of emergent \lya\ emission.  
The relevant angular scale for the optically-thick CGM \ion{H}{1} gas is $\simeq 10$\arcs\ ($\simeq 80$ physical kpc
at $z\sim 2.5$),  
whereas a typical extraction aperture for a slit spectrum is $\sim  1\secpoint2 \times 1\secpoint4$ -- a difference
of a factor of more than 180 in solid angle. Thus, even if the Case B-expected production rate of \lya\ photons were to 
escape the CGM of a galaxy, it is likely that the emission would be distributed over such a large region 
that a narrow slit would miss most of the \lya\ flux; even very deep narrow-band images
might leave much of the flux unaccounted-for due to limited surface brightness sensitivity.

In this paper, we present direct observational evidence showing that extended
\lya\ scattering ``halos'' are a generic property of high redshift star-forming galaxies, including
those that have no apparent \lya\ emission lines in their far-UV spectra. In \S2
we describe a sample of 92  UV-continuum-selected galaxies for which both rest-far-UV spectra
and deep narrow-band \lya\ images are available, and discuss the relationship between \lya\ 
properties measured using both techniques. 
In \S3 we use  
composite UV spectra, as well as \lya\ and continuum image stacking, to measure \lya\ 
emission extending to very low surface
brightness thresholds for various galaxy sub--samples. 
The results and their implications for the nature of \lya\ emission
in star-forming galaxies are described in \S4, discussed in \S5, and summarized in \S6.   

Throughout the paper we assume a Lambda-CDM cosmology with $\Omega_m = 0.3$, $\Omega_{\Lambda}=0.7$, and
$h=0.7$.  

\section{The Galaxy Sample}

The galaxies used in this paper are drawn from 3 survey regions where we used
UV-color selection to select galaxies with $1.8 \simlt z \simlt 3.4$ for spectroscopy (\citealt{steidel03,steidel04,shapley05}). 
In addition to completing
extensive ``Lyman Break Galaxy'' (LBG) spectroscopic follow-up, we have also imaged the 3 regions
using narrow-band (NB) filters centered at the observed wavelength 
of \lya\ at the redshift of galaxy over-densities we had previously identified from the
continuum-selected spectroscopic sample. 
Table~\ref{tab:obstable} summarizes the NB observations in these fields, all of which are among the deepest
NB images ever obtained for \lya\ at redshifts $z \sim 2-3$.     
The number of continuum-selected galaxies with spectroscopic redshifts falling within the redshift
range subtended by the NB filter bandpass in each field are also 
summarized in Table~\ref{tab:obstable}.

\begin{deluxetable*}{l c c c c c c c c c }
\tablewidth{0pt}
\tablecaption{Summary of \lya\ Narrow-Band Observations}             
\tablehead{
\colhead{Field}&
\colhead{NB\tablenotemark{a}}& 
\colhead{$z$\tablenotemark{b}}& 
\colhead{N\tablenotemark{c}}&
\colhead{Telescope/Instrument}&
\colhead{Date}&
\colhead{$t_{exp}$\tablenotemark{d}}&
\colhead{PSF\tablenotemark{e}}&
\colhead{$S(\lya,obs)$\tablenotemark{f}}&
\colhead{$S(\lya,z=2.65)$\tablenotemark{g}} 
}
\startdata
HS1549$+$195 &4667/88& $2.802-2.875$ & 27 & Keck 1/LRIS & 2010 May & 18,000 & $0\secpoint86$ & $1.29$ &  $1.59$  \\  
HS1700$+$643 &4018/90& $2.266-2.340$ & 43 & Palomar 5m/LFC & 2007 Jul & 80,280 & $1\secpoint20$ & $2.30$ & $1.54$ \\ 
SSA22a\tablenotemark{h} &4980/80 & $3.063-3.129$ & 22 & Keck 1/LRIS & 2005 Aug & 33,880 & $0\secpoint80$ & $0.94$ & $1.50$   \\
                          &   &   &    & Subaru/SuprimeCam & 2002 Sep & 25,800 &  & & 
\enddata
\tablenotetext{a}{Central wavelength/bandwidth of NB filter, in \AA. }
\tablenotetext{b}{Redshift range included between NB filter half-power points.}
\tablenotetext{c}{Number of continuum-selected, spectroscopically identified galaxies with NB measurements.}
\tablenotetext{d}{Total integration time, in seconds.}
\tablenotetext{e}{Stellar FWHM in arc seconds after smoothing to match the CB and NB PSF prior to photometry.}
\tablenotetext{f}{Observed surface brightness isophotal threshold (1.5$\sigma$), in units of 
$10^{-18}$ ergs s$^{-1}$ cm$^{-2}$ arcsec$^{-2}$.}
\tablenotetext{g}{Isophotal surface brightness threshold, corrected to $z=2.65$, in units of $10^{-18}$ ergs s$^{-1}$ cm$^{-2}$ arcsec$^{-2}$.}
\tablenotetext{h}{The field imaged with Keck/LRIS is a $5\minpoint5 \times 7\minpoint6$ subset of the LBG survey 
field from \cite{steidel03,steidel00}. The NB image is a combination of the LRIS images and archival Subaru images, discussed by Nestor et al 2011.}
\label{tab:obstable}
\end{deluxetable*}

We now briefly comment on each of the fields observed: 

\begin{itemize}

\item SSA22a has a redshift ``spike''
centered at $z = 3.09$ (\citealt{steidel98,steidel03}) which was first identified from the spectroscopic 
follow-up of LBGs. It was first imaged in \lya\ at the same redshift by \cite{steidel00}, who discovered
two very large ($>100$ kpc)``\lya\ Blobs'', prompting several subsequent studies of \lya-selected objects
using deeper narrow-band data (e.g., \citealt{matsuda04,hayashino04,nestor11}). 
Here we include the 22 continuum color-selected LBGs with spectroscopic
redshifts (\citealt{steidel03,shapley06}) lying within a $5\minpoint5 \times 7\minpoint6$ region with
especially deep \lya\ NB observations (Table~\ref{tab:obstable})

\item HS1700$+$64 is a survey field centered on the the position of
a hyper-luminous ($r'=16.0$, or $L \simeq 1.1\times10^{14} \lsun $) $z=2.751$ QSO. 
A galaxy over-density was again identified from spectroscopic follow-up,
with $z=2.299\pm0.03$ (\citealt{shapley05,steidel05}). We have subsequently obtained very deep NB imaging 
in both \Ha\ and \lya\ at this redshift (Erb et al 2011, in prep.). We include in the present sample 
the 43 continuum-selected  galaxies with spectroscopic redshifts placing the \lya\ transition
within a NB filter designed for follow-up of the proto-cluster. 

\item HS1549$+$195 is another survey field centered on the position of a hyper-luminous QSO ($r'=15.9$, or
$L \simeq 1.4 \times10^{14} \lsun $), with $z_Q=2.842$.
Once again a galaxy over-density was identified from the LBG spectroscopic follow-up, in this case centered
on the redshift of the QSO itself  
\footnote{ As we will show below, there is no evidence
that the presence of the QSO has significantly altered the overall \lya\ emission of the galaxies at the
same redshift.}. The NB4670 filter was designed to follow-up on the galaxy over-density, and in response to the serendipitous
discovery of spatially offset (and plausibly fluorescent) \lya\ emission associated 
with a $z=2.842$ damped \lya\ absorption (DLA) absorption system  
identified in the spectrum of a faint background QSO (\citealt{adelberger06}). 

\end{itemize}

In all 3 fields used in the present paper, galaxies were selected using rest-UV (LBG) color
selection and observed spectroscopically using the Keck 1 10m telescope and 
LRIS spectrograph (\citealt{oke95,steidel04}) {\it prior} to the \lya\ imaging,  
so the resulting sample should be relatively unbiased with respect to \lya\ properties. 
The full sample of 92 galaxies with mean redshift $\langle z \rangle = 2.65$ is broadly 
representative of UV-selected spectroscopic samples (e.g., \citealt{steidel03,shapley03,
steidel04,adelberger04})
in terms of both continuum and \lya\ properties: for example, they have  $23.4 \le {\cal R}_{AB} \le 25.5$
with median (mean) of $ {\cal R}_{AB}  =24.47$ (24.50), and   
spectroscopically-measured $W_0(\lya)$ in the range $-37$ \AA\ (absorption) to $+89$ \AA\ (emission) with  
median $W_0(\lya) \simeq +0.9$ \AA\ (cf. \citealt{shapley03,reddy08,kornei10}).  
The spectroscopic measures of \lya\ are based on extraction apertures 
of angular size $1\secpoint2$ (the slit width) by $\simeq 1\secpoint35$, independent of
wavelength, so that \lya\ and the UV continuum light are measured over identical spatial regions. 
We used a method similar to that described by \cite{kornei10} to measure $W_0(\lya)$ directly from the galaxy spectra.

\lya\ equivalent widths (and fluxes) were also measured for the same set of 92 galaxies using
a comparison of deep narrow-band (NB) and continuum (CB) images. As discussed by (e.g.) \cite{steidel00}, 
care must be taken since spectroscopic and imaging measurements of \lya\ may not be measuring the same
quantities.  
For measurements of \lya\ line emission from CB-NB color, the photometric aperture is
often defined by the region within an isophote corresponding to a particular \lya\ surface
brightness threshold, which of course depends on the depth of the \lya\ image.  
It also depends on the suitability of the continuum measurement for estimating the UV continuum 
flux density in the vicinity of the \lya\ line, which may require a color-term correction and/or correction
for \lya\ line contamination. 
The 3 \lya\ images used here are
comparably deep to the deepest \lya\ surveys to date, with 
1$\sigma$ surface brightness thresholds of
1.53, 0.86, and 0.63 $\times10^{-18}$ ergs s$^{-1}$ cm$^{-2}$ arcsec$^{-2}$ for the HS1700, HS1549, 
and SSA22 fields, respectively\footnote{Isophotal apertures corresponding to $1.5\sigma$ above the local
sky were used for NB-selected catalogs in all 3 cases.}. 
Although these observed SB thresholds differ by a factor of
more than 2, the deeper data at higher redshift result in rest-frame \lya\ surface brightness thresholds 
which differ by less than 10\%. The last column of Table~\ref{tab:obstable} shows the relative surface
brightness thresholds when all 3 datasets are shifted to the mean redshift of $\langle z \rangle = 2.65$. 

Important to generating \lya\ line images for relatively continuum-bright galaxies (and for
measuring line equivalent widths independently of spectroscopy) is a measure
of their far-UV continuum (hereinafter CB, or $m_{AB}[1220 \AA]$) near the wavelength of the \lya\ line.  
Ideally, the CB should
have the same effective wavelength as the NB without including the \lya\ line itself. For
the 3 fields presented here, deep CB images were created using 
linear combinations of two broad-band filters bracketing the NB passband; details of these
procedures are described in the individual papers cited above. 
Briefly, in HS1700 we used
very deep $U_n$(3550/700) and $G$(4730/1100) images obtained in 2001 May with the William Herschel 4.2m Telescope
prime focus imager (see \citealt{shapley05}) to create a ``UG'' continuum image with an effective wavelength of 4010\AA.
For SSA22a, we used archival B and V images taken with the 8.2m Subaru telescope with Suprimecam to create
a ``BV'' CB image with $\lambda_{\rm eff} = 4980$ \AA\ (see Nestor et al 2011). The HS1549 field
was treated somewhat differently, since the deepest broad band image (10,800 s integration with Keck/LRIS)
was obtained in the V band\footnote{Because the data were obtained using a dichroic with a transition wavelength of
$\simeq 5000$\AA\ (d500), the V passband was shifted to slightly longer wavelength ($\lambda_{\rm eff} \simeq 5506$ \AA\ instead
of 5464 \AA .} using the LRIS red channel contemporaneously with the March 2007 NB4670 images on the blue channel. 
A less-deep G-band image (2500 seconds with Keck/LRIS-B) was used to estimate the
appropriate (object-dependent) color correction needed to adjust the deeper V-band images to an effective wavelength near 4670 \AA. 
Since the observed range in continuum color among the sample galaxies at a given redshift is small (e.g., the mean and standard
deviation in observed broad-band color 
for the 43 $z \simeq 2.30$ galaxies in the HS1700 field is $\langle G-{\cal R} \rangle = 0.26\pm0.12$, and
$\langle U_n-G \rangle = 0.80\pm0.20$ where the standard deviation in both cases {\it includes}
photometric scatter), and the passbands are separated by only $\simeq 300$ \AA\ in the galaxy rest frame, we believe
that systematic errors associated with producing the CB image at the appropriate effective wavelength is likely very
small ($<< 0.1$ mag.).

In brief, the CB and NB images were first scaled to have matching zeropoints based on photometry of spectrophotometric
standard stars and by calculating the relative system throughput in each filter passband as a cross-check.  
The suitably scaled CB or NB 
images were smoothed using a Gaussian kernel to
match the stellar point-spread functions (PSFs) in the two images; the final PSF size for each field is listed
in Table~\ref{tab:obstable}. Matched aperture photometry was performed using dual
image mode in SExtractor (\citealt{bertin05}), with CB-NB colors measured using photometric apertures
defined in the NB image at an isophotal threshold equivalent to 1.5$\sigma$ per pixel above the estimated
local sky background. The CB zero point was iteratively adjusted by a small amount ($< 0.1$ mag in all cases) so that 
the median color of all objects in the image having $23 \le CB \le 26$ (the principle range expected
for the galaxies of interest) has CB-NB=0, corresponding to identical flux 
density measured in each band. The statistical error in the measurement of CB-NB can be conservatively estimated 
from the dispersion in color for all objects in the same range of apparent magnitude, which is quite small because
of the intrinsically narrow range in color and the depth of both the CB and NB images [$\sigma({\rm CB-NB}) 
\simeq 0.05-0.1$ mag.]    
A continuum-subtracted \lya\ line image (hereinafter ``\lya'' image)  
was formed by subtracting the scaled continuum image from 
the NB image 
\footnote{Among the 3 fields, only the SSA22a CB includes a small overlap ($\simeq 2.5$\% of its full bandwidth) with the NB \lya\ passband;
this would have the effect of a small (negligible for our purposes) over-subtraction of the continuum when producing the \lya\ line image.}.

We measured $W(\lya)$ (in units of \AA) from the CB-NB color using the simple relationship
\begin{eqnarray}
W_0(\lya) = B_{NB}~ \left[10^{0.4|CB-NB|}-1 \right]~ {CB-NB\over |CB-NB|} \AA\  
\end{eqnarray}
where $B_{NB}$ is the appropriate rest-frame bandwidth in \AA\ of the NB filter 
(19.6 \AA, 23.4 \AA, and 27.3 \AA\ for SSA22, HS1549, and HS1700, respectively). Note that
with this definition, $W_0(\lya) = 0$ when $CB=NB$, and positive (negative) values indicate net \lya\
emission (absorption). 
At an isophotal threshold of $\simeq 1.5 \times 10^{-18}$ ergs 
s$^{-1}$ cm$^{-2}$ arcsec$^{-2}$, the typical solid angle subtended by the detection isophote in the narrow-band (NB)
image for the continuum-selected  
galaxies is $\simeq 3.8$ arcsec$^{2}$, $\sim 2.3$ times larger than the spectroscopic aperture. 
If the spatial distribution of \lya\ emission is significantly different from that of the continuum light,
then the measured colors [and hence the inferred $W_0(\lya)$] could differ from the spectroscopic values. 
Figure~\ref{fig:ew_vs_ew} compares the measurements of $W_0(\lya)$ from the spectra versus those based on 
the CB-NB colors for the same 92 galaxies in the current sample, while Figure~\ref{fig:ew_hist} compares
the two distributions.  There is a modest tendency for 
the value of $W_0(\lya)$ measured from the NB images to be larger in absolute value (whether in absorption or emission)
near the extremes of the distribution, though they
have very similar mean and median values (Figure~\ref{fig:ew_hist}) and agree well when $|W_0|$ is small. 
If significant \lya\ flux were distributed on
still larger angular scales (with lower \lya\ surface brightness) while the same is not true of the UV continuum
light, then even the larger NB-based $W_0(\lya)$ would underestimate the true values.

\begin{deluxetable*}{lccccccccccc}[thb]
\tablewidth{0pt}
\tabletypesize{\scriptsize}
\tablecaption{Ly-$\alpha$ and Continuum Surface Brightness Profiles for Composites}
\tablehead{
\colhead{Sample\tablenotemark{a}} &
\colhead{Number} &
\colhead{$\langle z \rangle$} &
\colhead{$\langle m_{AB}\rangle$\tablenotemark{b}} & 
\colhead{$C_l$\tablenotemark{c}} &
\colhead{$b_l$\tablenotemark{c}} &
\colhead{$C_c$\tablenotemark{c}} &
\colhead{$b_c$\tablenotemark{c}} &
\colhead{$F(Ly\alpha)$\tablenotemark{d}} &
\colhead{$L_{tot}(Ly\alpha)$\tablenotemark{e}} & 
\colhead{$W_0(Ly\alpha,spec)$\tablenotemark{f}} &
\colhead{$W_0(Ly\alpha,tot)$\tablenotemark{g}} \\ 
\colhead{} &
\colhead{} &
\colhead{}   & \colhead{(\@1220 \AA)} & \colhead{($10^{-18}$)} & \colhead{(kpc)} & 
\colhead{($10^{-18}$)} & \colhead{(kpc)} & \colhead{($10^{-16}$)} & 
\colhead{($10^{42}$ ergs s$^{-1}$)} & \colhead{(\AA)} & \colhead{(\AA)} 
}
\startdata
All & 92 & 2.65 & 24.60 & ~2.4 & 25.2 & ~87.2 & 3.4 & ~1.7 & ~9.7 & $~+6.9$ & $+36.0$  \\
Ly$\alpha$ Em & 52 & 2.66 & 24.40 &  ~3.1 & 25.6 & 136.3 & 2.9 & ~2.5 & 14.3 & $+13.2$ & $+44.9$ \\ 
Ly$\alpha$ Abs & 40  & 2.63 & 24.72 & ~1.5 & 20.8 & ~52.5 & 4.5 & ~0.7 & ~4.0 & $~-4.4$ & $+16.8$  \\
All non-LAE & 74 & 2.65 & 24.56 & ~1.4 & 25.5 & 124.9 & 2.8 & ~1.4 & ~8.0 & $~+1.0$ & $+29.1$  \\
LAE only & 18 & 2.64 & 24.68 & ~3.9 & 28.4 & 110.3 & 2.9 & ~4.0 & 22.8  & $+29.2$ & $+92.9$  \\
Ly$\alpha$ Blobs & 11  & 2.59 & \nodata & 15.7 & 27.6 & \nodata & \nodata & 11.5 & 65.7 & \nodata & \nodata    
\enddata
\tablenotetext{a}{Galaxy sub-sample, drawn from the full sample (All) of 92 continuum-selected
galaxies with Ly$\alpha$ imaging. 
The details of the sub-samples are described in the text.}
\tablenotetext{b}{Average continuum apparent magnitude at $\lambda_0 \simeq 1220$ \AA, estimated from the CB photometry.}  
\tablenotetext{c}{Best fit parameters assuming SB profile $S(r) = C_n {\rm exp}(-b/b_n)$, where
$C_n$ is in units of $10^{-18}$ ergs s$^{-1}$ cm$^{-2}$ arcsec$^{-2}$. The sub-scripts
$l$ and $c$ refer to the Ly$\alpha$ line and UV continuum profiles, respectively.}
\tablenotetext{d}{Average integrated $\lya$ flux, in units of $10^{-16}$ ergs s$^{-1}$ cm$^{-2}$.}
\tablenotetext{e}{Average integrated $\lya$ luminosity, in units of $10^{42}$ ergs s$^{-1}$, assuming $\langle z \rangle = 2.65$.} 
\tablenotetext{f}{$\lya$ rest equivalent width measured from spectrum (Figure~\ref{fig:comp_spec}). }
\tablenotetext{g}{$\lya$ rest equivalent width of total Ly$\alpha$ flux, in \AA.} 
\label{table:tab2}
\end{deluxetable*}

\section{Inferences from Stacked Composites}

\subsection{Spectroscopic Stacks}

\begin{figure*}
\centerline{\includegraphics[width=19cm]{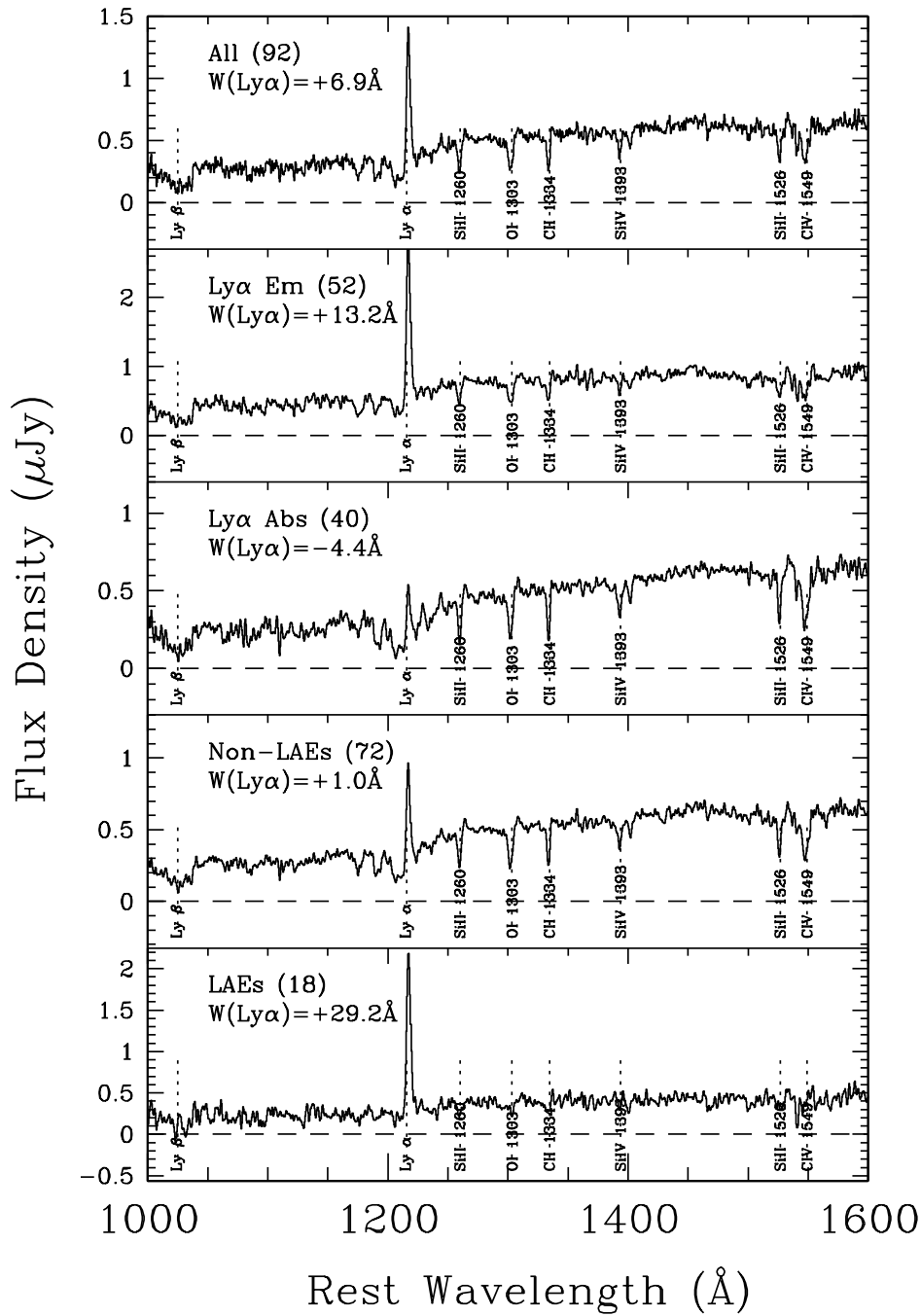}}
\caption{Composite spectra formed from the average within the subsamples
detailed in Table~\ref{table:tab2}. Within each panel, the number of galaxies going into the stack
is listed after the sub-sample
name; the second line in the annotation
lists the rest-frame equivalent width of the Ly$\alpha$ line measured from the composite
spectrum, with the convention that
positive values indicate net emission.  
\label{fig:comp_spec}
}
\end{figure*}

In order to measure \lya\ emission with SB well below the detection threshold for individual objects, 
we constructed composite spectra and images after dividing the sample of 92 into several subsets,
summarized in Table~\ref{table:tab2}. 
We used the values of $W_0(\lya)$ measured from the CB-NB colors for all galaxies,
with apertures defined by the isophotal thresholds listed in column 9 of Table~\ref{tab:obstable}. This method
has generally smaller statistical uncertainties and aperture corrections compared to the spectroscopic
measurements, and facilitates comparison with most deep \lya\ surveys, which are based primarily on equivalent
widths and fluxes inferred from the NB photometry 
\footnote{We have verified that none of the results of this paper depends significantly
on whether the imaging or spectroscopic measures of \lya\ are used to define the subsets.}.

As illustrated in Figure~\ref{fig:ew_hist}, the median $W_0(\lya)$ from both NB imaging and spectroscopic measurements 
is close to zero, in agreement with previous results for continuum-selected samples (\citealt{steidel00,shapley03,kornei10}).  
In forming subsets of the sample of 92, we used the NB \lya\ measurements   
to split the sample into ``\lya\ Em'', those that have \lya\ in net emission (52), and ``\lya\ Abs'', those having \lya\ in
net absorption (40)--see the second and third rows of Table~\ref{table:tab2}, respectively.  
Two other subsets were made consisting of galaxies satisfying the
criteria commonly adopted for ``Lyman $\alpha$ Emitters'' (LAEs), i.e., $W_0(\lya) \ge 20$ \AA,
of which there are 18 (20\% of the total), with the remainder (74 of 92, or 80\%) 
placed in a sub-sample called ``non-LAEs'', i.e., all continuum-selected LBGs that would not be selected as LAEs. 

For each sub-sample
listed in Table~\ref{table:tab2}, a composite far-UV spectrum was created
by shifting the observed, flux-calibrated spectra into
the galaxy rest-frame using the prescriptions given in S2010, re-sampling the rest-frame spectra to $0.5$ \AA\ pix$^{-1}$, 
and averaging. Each stacked composite spectrum was scaled so that the continuum
level near \lya\ matched that obtained from the photometric stack of the same subset of galaxies, discussed
below. 
The correction was typically a factor of $\simeq 1.5$, and was applied for the sole purpose
of placing the continuum levels for the same subsets on the same flux scale. 
The resulting stacked spectra are shown 
in Figure~\ref{fig:comp_spec}; measured properties of the composite spectra 
are given in the figure and listed in Table~\ref{table:tab2}.

\subsection{\lya\ and CB Stacks}

\begin{figure*}[htb]
\centerline{\includegraphics[width=8cm]{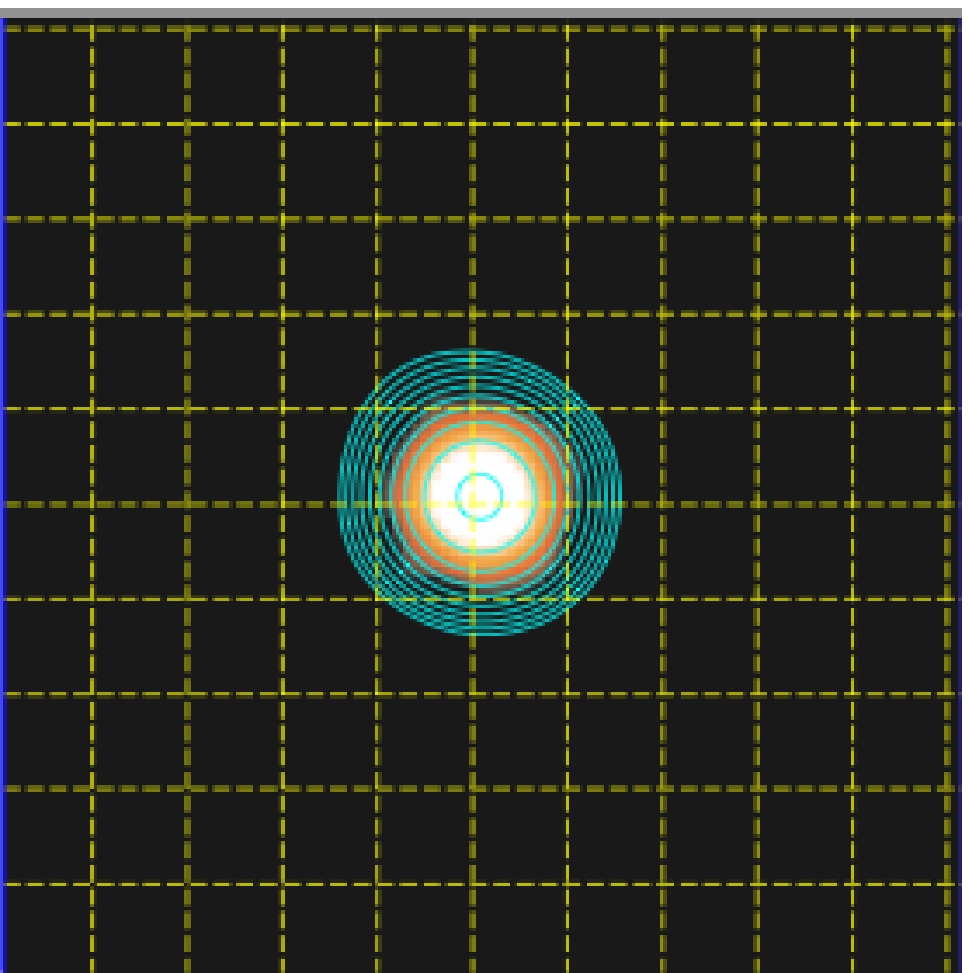}\includegraphics[width=8cm]{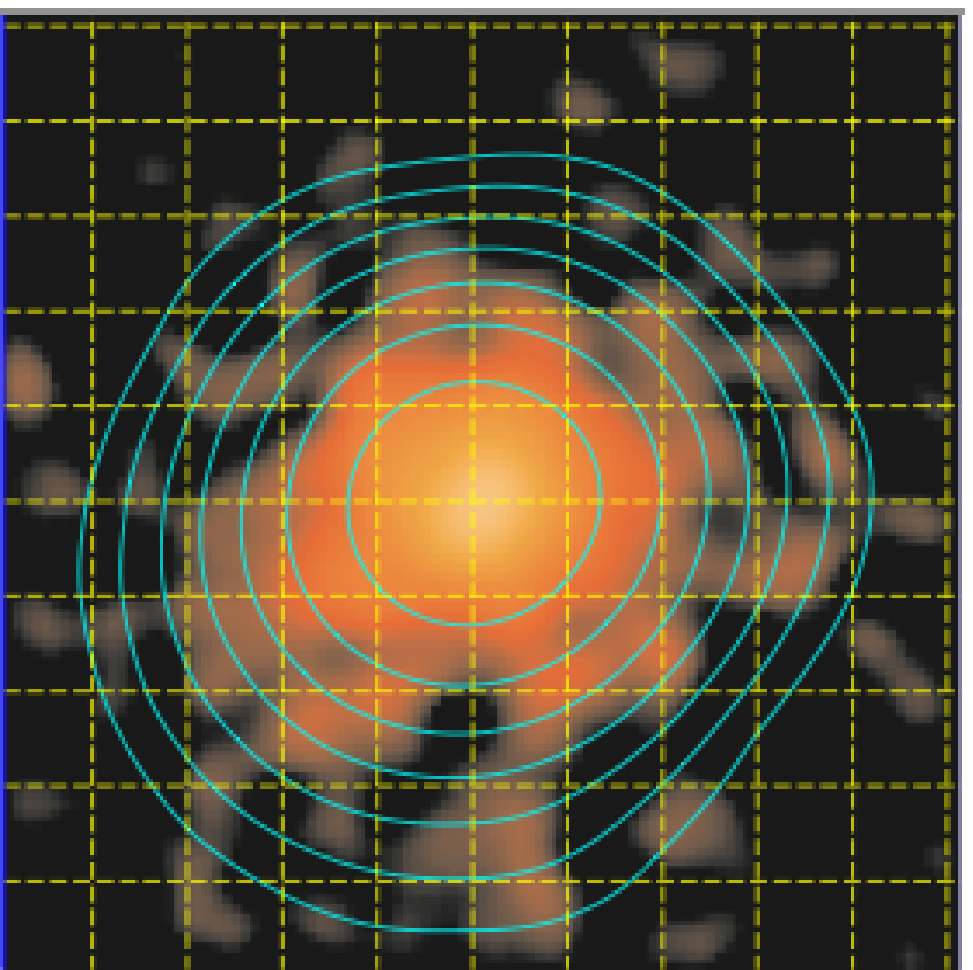}}
\caption{{(Left:)} Scaled far-UV continuum image produced (as described in the text) from the 
average of 92 continuum-selected LBGs, drawn from 3 independent fields. The regions shown are 20\arcs\
($\simeq 160$ kpc physical at $z=2.65$) on a side, with a grid spacing of 2\arcs.   
{(\it Right:)} The continuum-subtracted, stacked \lya\ image for the same sample of galaxies. In both
panels, the contours are logarithmically spaced in surface brightness 
with the lowest contour shown at $\simeq 2.5\times 10^{-19}$ 
ergs s$^{-1}$ cm$^{-2}$ arcsec$^{-2}$. 
\label{fig:Lya_stack}
}
\end{figure*}

\begin{figure}[thb]
\centerline{\includegraphics[width=9cm]{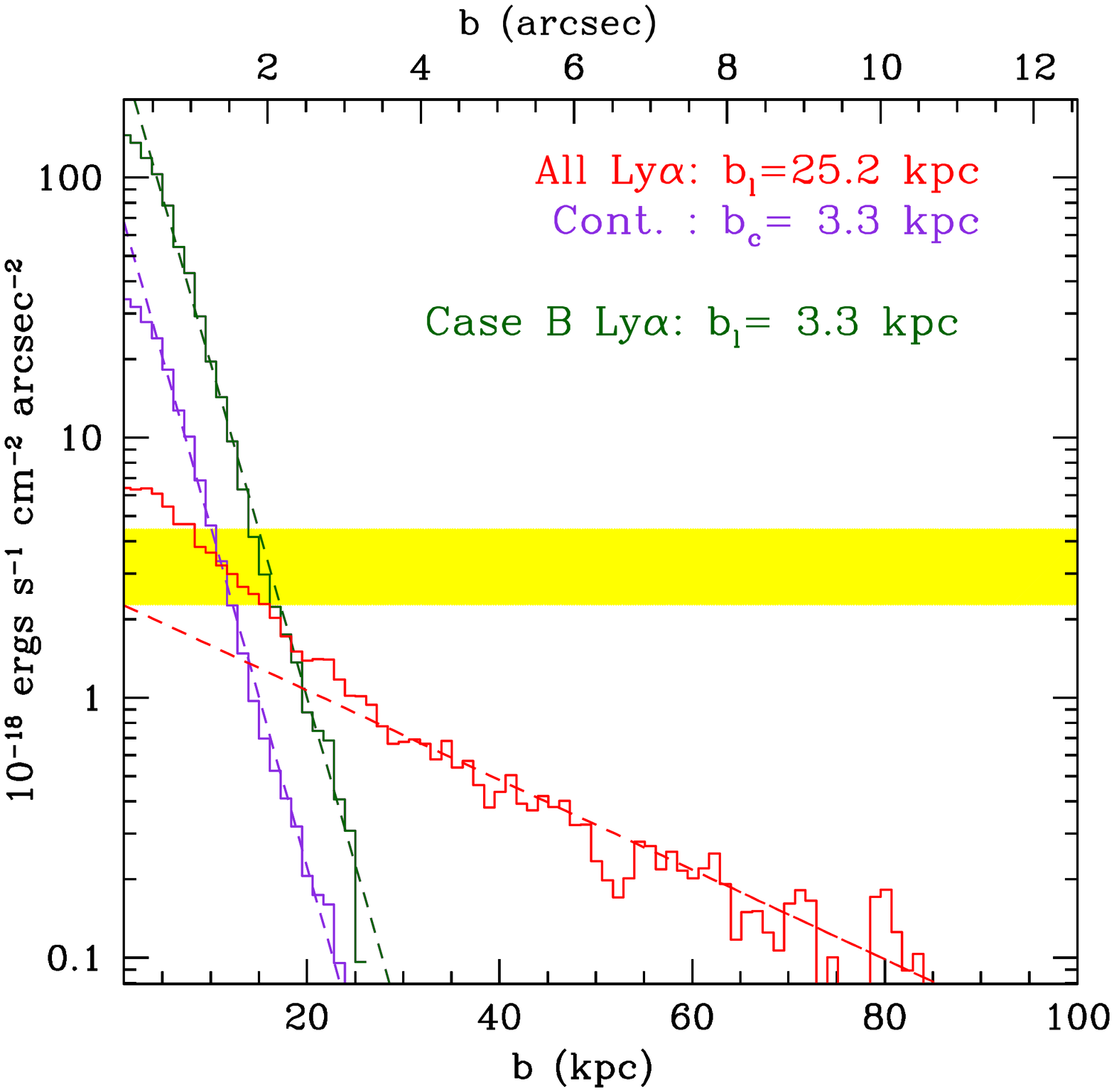}}
\caption{The observed average surface brightness profile for the 1220 \AA\ continuum
light (blue) and the \lya\ line (red) for the full sample of 92 continuum-selected galaxies, 
evaluated over the same rest-frame bandwidth sampled by the \lya\ image (24.3 \AA). 
Note that these profiles are simply the azimuthal averages of the stacked images shown 
in Figure~\ref{fig:Lya_stack}.  
The light-shaded region indicates the range of typical \lya\ surface brightness threshold reached by 
deep \lya\ surveys for the detection of individual objects.  
The dashed lines show the surface brightness profile assuming
that $S(b) = C_l{\rm exp}(-b/b_l)$ with parameters given in Table~\ref{table:tab2}.
The corresponding angular scale at $\langle z \rangle = 2.65$ is given
along the top axis. For the purpose of comparison, we also show the \lya\ profile 
expected for the same sources under the assumption of ``Case B'' \lya\ to CB ratio, no
destruction of \lya\ by dust, and no spatial diffusion of \lya\ photons due to resonant scattering
(i.e.,  the \lya\ and CB profiles would be identical in shape, and since $W_0(\lya)=100$ \AA,
the \lya\ line image would be a factor
of $\simeq 4.1$ brighter than the continuum in the effective rest-frame bandwidth of 24.3 \AA.)  
\label{fig:sb_profile}
 }
\end{figure}

A 25\arcs\ x 25\arcs sub-image (``postage stamp'') 
centered on the position of the continuum centroid of each
galaxy was extracted from the CB image and the (continuum-subtracted) 
\lya\ images after scaling them to a common zero point as discussed above. 
Masks were created by performing object detection on each continuum sub-image 
using SExtractor (\citealt{bertin05}). These were used to exclude pixels 
lying within the detection isophotes of any   
object other than the central one, and were applied during the stacking to both the CB and \lya\ images.   
Two stacked images were formed for each subset listed in Table~\ref{table:tab2} (straight averages, with masking),
one for the CB image and another 
for the \lya\ line image. 
Figure~\ref{fig:Lya_stack} compares the average CB image with the \lya\ image for the full 
sample of 92 galaxies, while Figure~\ref{fig:sb_profile} compares the azimuthally-averaged
surface brightness profiles of the same composite CB and \lya\ images. 
Figure~\ref{fig:Lya_stack} shows clearly that, on average, 
\lya\ emission
is detected to radii of at least 10\arcs, or $\simeq 80$ physical kpc at $\langle z \rangle = 2.65$. Figure~\ref{fig:sb_profile} 
shows that the average CB light profile for the same galaxies is much more compact and drops
below the SB detection threshold for $b \simgt 20$ kpc ($\simgt 2\secpoint5$).  
Figure~\ref{fig:sb_profile} also shows what the \lya\ line profile would look like if $W_0(\lya) = 100 \AA$ (i.e., Case B)
and \lya\ and CB light had the same spatial distribution on average.

The surface brightness profiles for both the continuum and the \lya\ line images are reasonably well-fit
by an exponential of the form $S(b) = C_i{\rm exp}(-b/b_i)$ for projected radii beyond the central arcsec; 
the parameters of the best-fit values for the normalization $C_i$ and scale length $b_i$ 
are given in Table~\ref{table:tab2} for
each sub-sample as well. In the full image stack, the effective surface brightness detection thresholds 
are a factor of $\sim 10$ lower than for individual galaxies; it is clear that the distribution
of \lya\ emission is very different from that of the continuum for every sub-sample, with
best-fit \lya\ scale lengths of $b_l \simeq 20-30$ kpc compared to the corresponding continuum
emission which has $b_c \simeq 3-4$ kpc. It is important to note that the true difference in scale length
is larger, since we have made no attempt to de-convolve the profiles from
the seeing disk, which was FWHM$\simeq$ 0.86, 1.20, and 0.80\arcs\ for HS1549, HS1700, and SSA22a, respectively.
The continuum profiles of the stacked composite CB images have $FWHM \simeq 1\secpoint2-1\secpoint4$,
indicating average (seeing de-convolved) galaxy continuum
sizes of FWHM$\simeq 0.80$\arcs\ ($\sigma \simeq 0.35$\arcs). These de-convolved angular sizes are also
consistent with measurements of similar galaxies in deep HST/ACS images (e.g., \citealt{peter07,law07}.)

\begin{figure}[thb]
\centerline{\includegraphics[width=9cm]{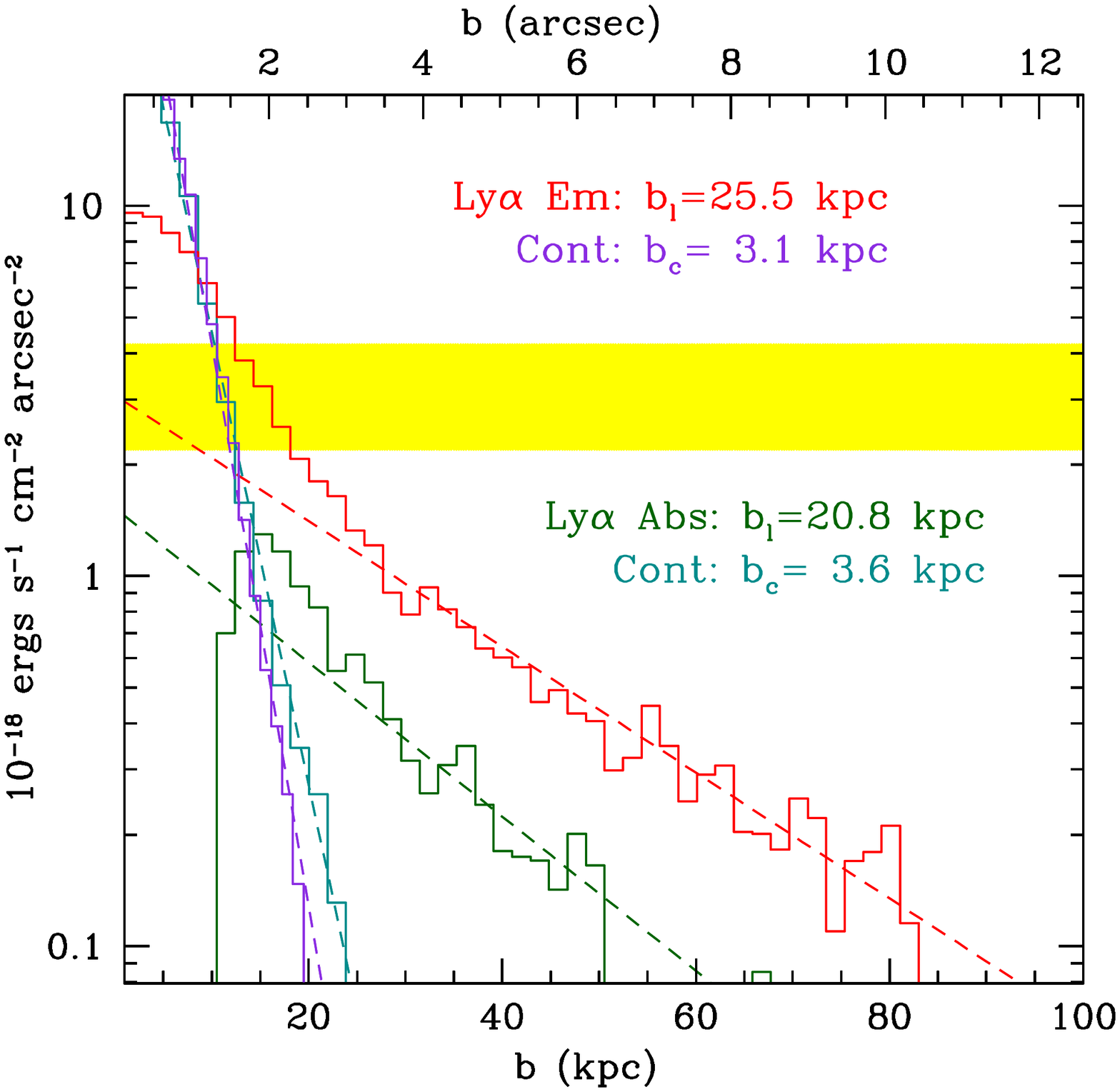}}
\caption{Same as Figure~\ref{fig:sb_profile}, comparing the average surface
brightness profiles for the the sample divided according to whether the NB measurements
indicate net \lya\ ``Abs'' or ``Em''.  
\label{fig:emabs_plot}
} 
\end{figure}

\begin{figure}[thb]
\centerline{\includegraphics[width=9cm]{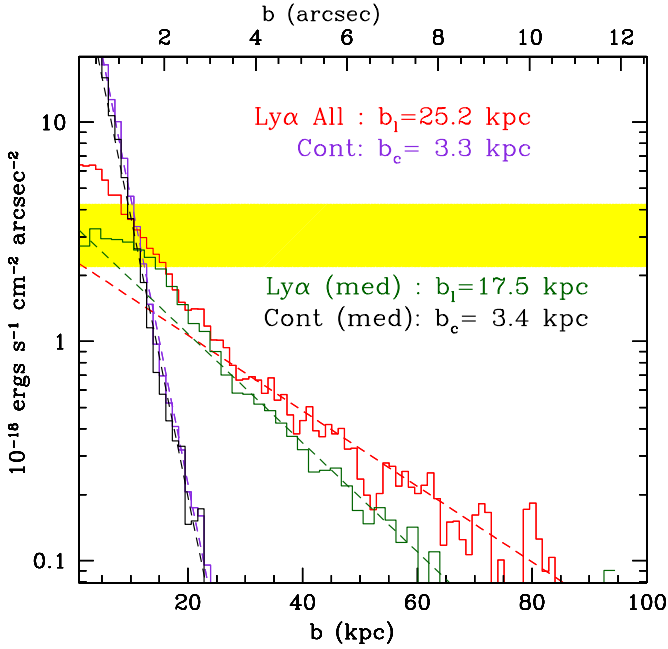}}
\caption{Same as Figure~\ref{fig:sb_profile}, comparing the surface brightness profile
measured from median-combined rather than mean-combined stacks of \lya\ and 
CB sub-images. Note that the line-to-continuum ratio is not necessarily preserved in
the median stacks. The general effect of this alternative processing is to decrease 
the measured \lya\ scale lengths by $\sim 25$\%, though the \lya\ profiles remain
much more extended than the continuum profiles.
\label{fig:plot_avg_med}
}
\end{figure}

The stacked \lya\ and CB images as in Figures~\ref{fig:sb_profile} and \ref{fig:emabs_plot} represent unweighted averages
of all galaxies in the sample (with masking as described above). This choice was motivated by the desire 
to preserve the photometric integrity of the stacks so that fluxes could be measured directly using aperture
photometry, but also because any scaling or weighting would require deciding what the relevant figure of merit
should be. Medians are often used to suppress outliers in stacked data sets, but they have the disadvantage
for the present application of not preserving flux in two-dimensional images, of working best when scaling has been applied to individual
images going into the stack, and of suppressing real signal as it approaches the noise level. Nevertheless, in
Figure~\ref{fig:plot_avg_med} we show a comparison of the surface brightness profiles for median-combined
stacks as compared to mean-combined for both line and continuum.

\begin{figure}[thb]
\centerline{\includegraphics[width=9cm]{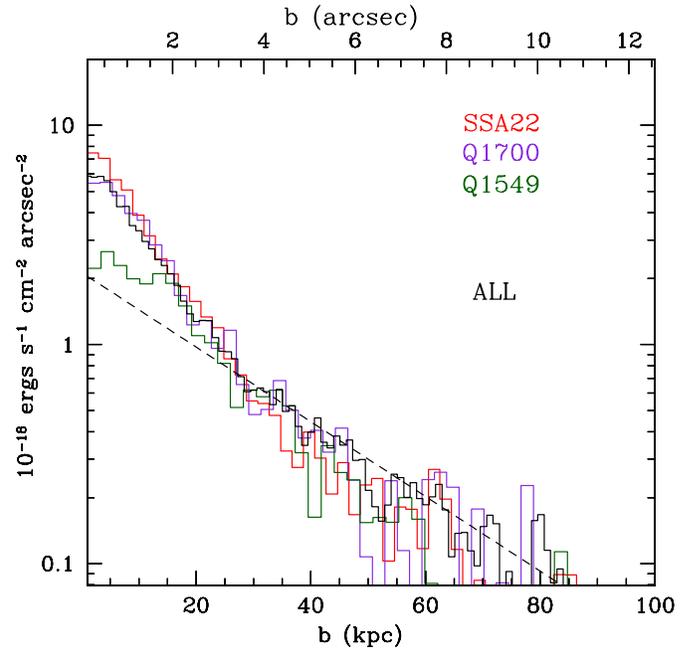}}
\caption{Average observed \lya\ surface brightness profiles for galaxy sub-samples separated
by field, as indicated. The \lya\ profiles were scaled according to the relative continuum
flux density in each field, for display purposes. 
The small differences
in the mapping of angular scale to physical scale have also been removed to facilitate the comparison.
The measured values of the average \lya\ rest equivalent
width are ${\rm \langle W_0(\lya) \rangle =28.1}$ \AA\, 28.6 \AA, and 42.0 \AA\ for HS1549, HS1700, and SSA22, respectively. 
\label{fig:plot_3fields} 
}
\end{figure}

We have argued that our sample of galaxies with $\langle z \rangle =2.65$ has emission line and continuum
properties characteristic of those in the full LBG spectroscopic surveys at these redshifts.  
Figure~\ref{fig:plot_3fields} shows that the diffuse \lya\ emission is also consistent among the 3 survey fields taken individually. 
This is important, since each field samples galaxies at a different redshift with observations subject
to a different set of conditions, using different telescope and instrument combinations.  
We also note that the bright QSO known to lie within the survey volume in the HS1549 field appears
not to have had a significant effect on the \lya\ emission from the galaxies in our sample\footnote{The galaxy regions most
likely to be affected by excess ionizing radiation from the QSO would lie in the outer parts; whether this radiation would
increase or decrease the amount of \lya\ emission from galaxies would depend on the physical state of the gas. The galaxies 
in the HS1549 field are fainter by about 30\% on average (in terms of apparent continuum magnitude) than in the other two fields;
the differences in the \lya\ profiles on small scales may be a consequence of this selection issue.}.

Figure~\ref{fig:emabs_plot} shows the measured surface brightness profiles and best-fit exponential
parameters for the sub-sample with net \lya\ emission (see the second row of Table~\ref{table:tab2}).
The profile of the \lya Em composite is qualitatively similar to that of the full sample. The main difference is in the
central \lya\ surface brightness --- 
\lya\ Em objects have (on average) \lya\ surface brightness well above
the threshold for individual detection (light shaded region in Figure.~\ref{fig:sb_profile} and \ref{fig:emabs_plot})
to projected distances of $b \sim 20-25$ kpc (2\secpoint5$-$3\secpoint1),
whereas even the peak \lya\ SB in the full sample (Figure~\ref{fig:sb_profile}) barely reaches the
(individual) detectability threshold. 
Also illustrated in Figure~\ref{fig:emabs_plot} is the SB profile for the ``\lya\ Abs'' sub-sample. 
The \lya\ SB scale length for the Abs sample is still $\simeq 4$ times
larger than for the corresponding continuum light (third row of Table~\ref{table:tab2})-- and has a comparable
\lya\ scale length $b_l$ to that measured for the \lya\ Em sample (second row of Table~\ref{table:tab2}),
albeit with a $\simeq 3$ times lower normalization for $b \simgt 20$ kpc. Clearly the difference
is much larger for $b \simlt$1\arcs, where the Abs sample exhibits a large ``hole'' in which \lya\ absorption
strongly dominates.    

\begin{figure}[thb]
\centerline{\includegraphics[width=9cm]{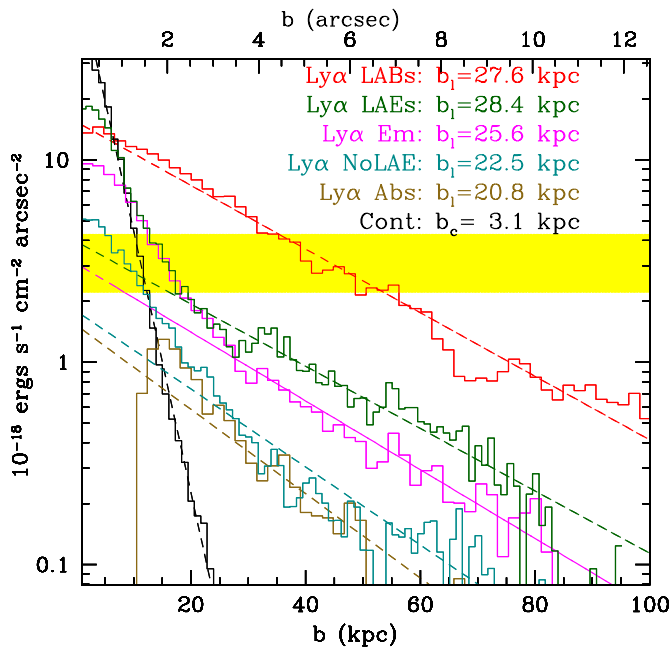}}
\caption{As for Figure~\ref{fig:sb_profile}, where here the \lya\ radial SB profiles are
shown for each of the sub-samples in Table~\ref{table:tab2},
along with the exponential models for each.  
Also included for comparison is the average surface brightness profile of 11 giant \lya\ ``Blobs'' observed
in the same 3 survey fields (red). 
\label{fig:blob_plot}
} 
\end{figure}

Figure~\ref{fig:blob_plot} reproduces the SB profiles of the ``\lya\ Em'' and ``\lya\ Abs'' sub-samples 
together with the average profile of LAEs (green), and non-LAEs (cyan). 
The LAE 
sub-sample is very similar to that of the larger ``\lya\ Em'' subset, but has an average SB a factor
of $\sim 1.5$ higher for $b \simgt 10$ kpc (and a factor $\simeq 3$ higher for $b \le 10$ kpc.) 
Also plotted for comparison (Figure~\ref{fig:blob_plot})
is the average SB profile of 11 ``\lya\ Blobs'' (LABs; see e.g. \citealt{steidel00,matsuda04}), which
for the present we define as \lya\ selected objects with detected isophotal diameters
$d > 5$\arcs, discovered in the same 3 survey fields. None of the 11 Blobs is included in
the main galaxy sample, since they do not have central continuum sources that satisfy the usual
LBG color criteria. As indicated in Table~\ref{table:tab2}, the Blobs have an average \lya\ luminosity
$\simeq 7$ times higher than an average galaxy in our sample.   
It appears that even the most extreme LABs do not have fundamentally different 
SB profiles compared to those of typical galaxies in the sample
{\it except} that their surface brightness normalization exceeds the typical detection threshold to $b \sim 50$ kpc ($\sim 6$\arcs).
In other words, {\it if one were routinely sensitive to a surface brightness of $\sim 10^{-19}$ ergs s$^{-1}$ cm$^{-2}$
arcsec$^{-2}$, all continuum-selected LBGs would be ``Blobs''}\footnote{ Conversely, if the LABs were several times
less luminous but had the same surface brightness profile, they would fail to be recognized as ``blobs'' at all.}. 
We will return to a more detailed discussion of \lya\ demographics in \S4.  

\begin{figure}[thb]
\centerline{\includegraphics[width=9cm]{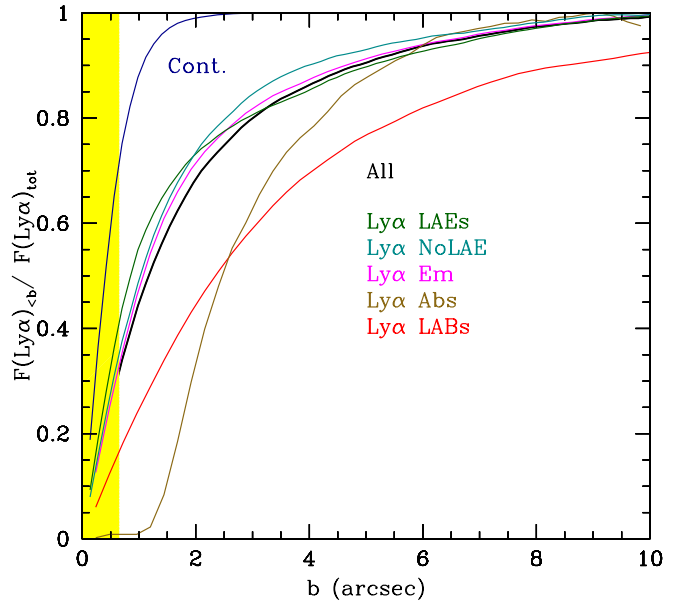}}
\caption{The cumulative fraction of the large-aperture \lya\ flux as a function of angular aperture radius $b$ 
for each galaxy sub-sample. The vertical
dashed line drawn at $b = 0\secpoint65$ indicates the typical effective aperture for the slit spectra of the same objects. The
dark blue curve corresponds to the cumulative continuum flux (for the stack of the full sample) as a comparison. 
\label{fig:lya_cum_plot}
} 
\end{figure}

If we assume for the moment that the extended \lya\ halos represent photons 
originating in the galaxies' \ion{H}{2} regions, we can use the composite 
\lya\ line and CB images to measure the integrated \lya\ line-to-continuum ratio, usually
parametrized as $W_0(\lya)$, the \lya\ equivalent width. The total \lya\ fluxes have been measured
directly from the calibrated stacked images (column 8 of Table~\ref{table:tab2}); a comparison with the continuum flux density
measured near the wavelength of \lya\ from the CB images (column 3 of Table~\ref{table:tab2}) allows the calculation of $W_0(\lya,tot)$ (column 12).   
These numbers can be compared directly with the spectroscopic measurements (column 11) for the same galaxy sub-samples.
The values that include the diffuse \lya\ extending to $\sim 80$ kpc radii around galaxies exceed the spectroscopically-inferred
$W_0(\lya,spec)$ by an average factor of $\simeq 5$ for the full galaxy sample.  
Figure~\ref{fig:lya_cum_plot} shows the cumulative fraction of the total \lya\ flux as a function of aperture radius $b$
(in arc seconds) for the galaxy samples in Table~\ref{table:tab2}.  
It is interesting to note that including the spatially extended \lya\ emission brings the average galaxy into
the range that would nominally qualify as a LAE ($W_0(\lya) > 20$ \AA)-- even for the ``No LAE'' sub-sample
that explicitly {\it excludes} the 18 conventional LAEs (row 4 of Table~\ref{table:tab2}). 

One can also compare the measured large-aperture $W_0(\lya)$ with expectations for the \lya\ to continuum
ratio for Case-B recombination
and no dust. As discussed above, an asymptotic value of $W_0(\lya) \simeq 100$ \AA\ is expected
when star formation has been continuous for $t_{SF} \simgt 4\times 10^7$ years\footnote{The typical inferred 
age for galaxies similar to those in the present sample is $\sim 500$ Myr (e.g., \citealt{shapley05,reddy08}).}.  
Note that approximately the same value of $W_0(\lya)$ is expected as long as \lya\ photons do not
suffer greater attenuation by dust than continuum photons just off the \lya\ resonance. Thus, the fact
that most of the values of $f_{esc,rel} \equiv W_0(\lya)/100$\AA\  (Table~\ref{table:tab3}) are significantly 
smaller than unity means that \lya\
photons suffer greater extinction than the continuum, by factors ranging from $1.1 - 6.0$ with an 
average $\simeq 2.8$ for the full sample of 92 galaxies. 

We note that the values in Table~\ref{table:tab3} for $f_{esc,rel}$ have been obtained using 
a method that appears to differ from that used in some recent work (e.g., \citealt{gronwall07,
nilsson09,kornei10}). Most estimates of $f_{esc,rel}$ use stellar population synthesis models to estimate
the level of extinction, which is then used to derive SFR to calculate the expected \lya\ luminosity
based on the assumption of Case B recombination and the form of the stellar IMF. While we are using largely
identical SED modeling
to estimate $SFR_{UV,corr}$, we use the observed $W_0(\lya)$ as a direct observational estimate of $f_{esc,rel}$. 
The present method relies on the same assumptions about the 
stellar IMF (i.e., based on a Salpeter-like IMF for high mass stars) 
to estimate the \lya\ photon production rate per unit star formation; the difference is that we rely
on the ratio of \lya\ photon production to that of 
$\lambda_0\simeq 1220$ \AA\ continuum photons from
the same ensemble of stars.  The advantage
of using the equivalent width measurement is that it should be independent of extinction if  
\lya\ and $\simeq 1220$ \AA\ continuum photons experience the same attenuation, and would directly reflect
the relative attenuation of line and continuum if $A_{\lya} \neq A_{1220}$. Because both of these methods
rely on measuring the integrated \lya\ line flux, an underestimate of \lya\ relative to the continuum
will cause $f_{esc,rel}$ to be underestimated by the same factor. For the present purposes, we
prefer using the method relying on $W_0(\lya)$ since it depends on a largely independent measurement that
may avoid propagating possibly large systematic errors in the estimates of E(B-V) from SED fitting 
[or from the assumed extinction curve, which relates E(B-V) to $A(\lambda)$]
to the calculation of $f_{esc,rel}$. In general, we expect that the under-counting of \lya\ photons due to the 
aperture effects discussed above are likely to dominate any differences in inferred $f_{esc,rel}$.

Column 10 of Table~\ref{table:tab2} compiles the average \lya\ luminosity $L_{tot}(\lya)$ implied by the
measured value of $F_{tot}(\lya)$ assuming the sample mean redshift $\langle z \rangle = 2.65$.
If one naively converts these numbers to an equivalent star formation rate (i.e., divide
by $2 \times 10^{42}$ ergs s$^{-1}$ to yield SFR in units of $M_{\sun}$ yr$^{-1}$) the results range
from $2.1 -11.8$ M$_{\sun}$ yr$^{-1}$ for the various sub-samples, with an average of 3.1 M$_{\sun}$ yr$^{-1}$
for the full sample.  In Table~\ref{table:tab3} we have compiled the statistics of the far-UV inferred SFRs and
continuum extinction 
for each of the sub-samples from Table~\ref{table:tab2}. 
The extinction estimates are parametrized by $E(B-V)$ and assume the \cite{calzetti00} starburst
attenuation curve; E(B-V) was estimated from SED fits when available,
and using the far-UV continuum slope for the $\sim$20\% of galaxies lacking adequate near-IR photometric coverage
for SED fitting. In the context of the assumed 
starburst attenuation relation, the extinction (in magnitudes) at 1500\AA\ is 
$A(1500 \AA) = 11.16 E(B-V)$ (\citealt{meurer99,calzetti00,reddy06,reddy10}).  
The UV continuum magnitudes near rest-frame 1500 \AA\ were used to estimate $SFR_{UV}$ (e.g., \citealt{mpd98,steidel99,as2000})
with a median value of $\langle SFR_{UV} \rangle \simeq 6.3$ M$_{\sun}$ yr$^{-1}$. Applying the median E(B-V) to the 
median $SFR_{UV}$ within each sub-sample implies that $ 11.3 \simlt \langle SFR_{UV,corr} \rangle \simlt 46.5$ 
M$_{\sun}$ yr$^{-1}$, with an overall median of $SFR_{UV,corr} = 34.3$ M$_{\sun}$ yr$^{-1}$--- 
very close to the mean of the LBG sample observed in \Ha\ by \cite{erb+06b} and consistent with the mean
bolometric luminosity of identically-selected LBGs estimated using multiple SFR indicators \citep{reddy06,reddy10}.    
The median $E(B-V)$ varies considerably among the sub-samples, so that the median attenuation of the UV
continuum is inferred to range from $\simeq 2.5$ for the LAEs to $\simeq 7$ for the \lya\ Abs sub-sample. In
Table~\ref{table:tab3} we list the inverse of this factor, which we have called $f_{esc}(UV) \equiv {\rm SFR_{UV}/
SFR_{UV,corr}}$. Our estimate of the fraction of all \lya\ photons produced by photoionization in the
galaxy \ion{H}{2} regions that have been detected is then given by  $f_{esc,tot}(\lya) \equiv f_{esc,rel}(\lya)\times f_{esc}(UV)$. These values
range from $f_{esc, tot}(\lya) \simeq 0.37$ for the LAE sub-sample to $f_{esc,tot}(\lya) \simeq 0.024$ for 
the ``\lya\ Abs'' sub-sample. The average for the entire sample is  
$f_{esc,tot}(\lya) \simeq 0.061$. We note that this fraction is close to the average
value of $f_{esc}(\lya)$ estimated by \cite{hayes10} based on a very different approach involving a comparison of
\Ha\ and \lya\ luminosity density at $z \simeq 2.2$.  

The last column of Table~\ref{table:tab3} shows the inferred ratio $A(\lya)/E(B-V)$, where both quantities
are expressed in magnitudes and $E(B-V)$ is inferred from the stellar SED. 
The numeric value of this ratio is $\simeq 19.5\pm1.5$ for all sub-samples except the LAEs,
which have $A(\lya)/E(B-V) \simeq 11.9$.  Since the \cite{calzetti00} extinction curve predicts that the 
UV continuum near the \lya\ line has $A(1220\AA)/E(B-V) \simeq 12$, it seems that \lya\ emission
from the LAEs drawn from our LBG sample exhibit no evidence for selective extinction of \lya\ photons, while
for other LBGs the attenuation $A(\lya)$ is $\sim 1.6$ times higher than for continuum photons for
the same value of $E(B-V)$. If the \lya\ escape fraction is controlled by processes confined to \ion{H}{2} regions, 
the result suggests that $E(B-V)_{neb} \simeq \eta~  E(B-V)_{stars}$ with $\eta ~\simeq 1.6$, on average. This   
can be compared 
with the relationship inferred for nearby star-forming galaxies, $\eta~ \simeq 2.5$, based
on measurements of the Balmer decrement \citep{calzetti00}.  
At present, there are few galaxies for which \lya\, \Ha, and \Hb\ have all been measured, though 
there are some indications that the same value of $E(B-V)$ applies for both continuum starlight and \Ha\ 
for galaxies similar to those in the current 
sample (\citealt{erb+06c}; but see \citealt{forst09} for possibly conflicting evidence).  

\begin{deluxetable*}{lccllcccc}
\tablewidth{7.0in}
\tablecaption{Inferred Continuum and Ly$\alpha$ Attenuation for Sub-Samples}
\tablehead{
\colhead{Sample\tablenotemark{a}} &
\colhead{$\langle m_{AB}(1500)\rangle$\tablenotemark{b}} &
\colhead{${\rm E(B-V)}$\tablenotemark{c}} &
\colhead{${\rm SFR_{UV}}$\tablenotemark{d}} &
\colhead{${\rm SFR_{UV,corr}}$\tablenotemark{e}} &
\colhead{${\rm f_{esc}(UV)}$\tablenotemark{f}} & 
\colhead{${\rm f_{esc,rel}(\lya)}$\tablenotemark{g}} &
\colhead{${\rm f_{esc,tot}(\lya)}$\tablenotemark{h}} &
\colhead{${\rm A(\lya)/E(B-V)}$\tablenotemark{i}}  
}
\startdata
All & $24.47/24.50\pm0.55$ & 0.17 & ~6.3 & 34.3 & 0.17 & 0.36 & 0.061 & 17.9 \\  
Ly$\alpha$ Em & $24.50/24.55\pm0.56$ & 0.11 &  ~6.0 & 18.6 & 0.32 & 0.45 & 0.144 & 19.1 \\ 
Ly$\alpha$ Abs & $24.44/24.43\pm0.54$ & 0.19 & ~6.6 & 46.5 & 0.14 & 0.17 & 0.024 & 21.3 \\ 
All non-LAE  & $24.43/24.42\pm 0.54$ & 0.18 & ~6.7 & 42.6 & 0.16 & 0.29 & 0.046 & 18.5 \\ 
LAE only  & $24.85/24.80\pm0.50$ & 0.09 & ~4.5 & 11.3 & 0.40 & 0.93 & 0.372 & 11.9 
\enddata
\tablenotetext{a}{Galaxy sub-sample, drawn from the full sample (All) of 92 continuum-selected
galaxies with Ly$\alpha$ imaging. 
The details of the sub-samples are described in the text.}
\tablenotetext{b}{Median and mean/standard deviation of continuum apparent magnitude at $\lambda_0 \simeq 1500$ \AA}  
\tablenotetext{c}{Median ${\rm E(B-V)}$ inferred from SED fitting.}
\tablenotetext{d}{Median SFR, in M$_{\sun}$ yr$^{-1}$, from UV continuum with no dust correction.}
\tablenotetext{e}{Median SFR after correction based on E(B-V) and Calzetti (2000) reddening relation.} 
\tablenotetext{f}{Fraction of 1500 \AA\ photons escaping galaxy.}
\tablenotetext{g}{Relative escape fraction of \lya\ photons, $\equiv {\rm W_0(\lya)_{tot}/100}$ \AA.}
\tablenotetext{h}{Fraction of \lya\ photons escaping, ${\rm f_{esc,tot} = f_{esc,rel}\times f_{esc}(UV)}$.}
\tablenotetext{i}{Ratio of attenuation of \lya\ photons to ${\rm E(B-V)}$ when both are expressed in magnitudes.}
\label{table:tab3}
\end{deluxetable*}

\section{Implications of Diffuse \lya\ Halos}

Diffuse \lya\ emission from the outer parts of actively star forming 
galaxies is an unavoidable consequence of 
a gaseous CGM so long as some component of it is optically thick to \lya\ photons and some
fraction of \lya\ photons initially produced in \ion{H}{2} regions are not absorbed by dust  
at smaller galactocentric radii.  
Calculation of the emergent \lya\ emission is undoubtedly complex, since it will depend
on the details of the gas-phase structure and kinematics as well as the relative distribution 
of the sources (e.g., \ion{H}{2} regions) and the sinks (e.g., dust) 
of \lya\ photons. 
A fully successful model requires 3-D radiative transfer calculations and all of the relevant spatial
and kinematic information as input. Such a treatment is far beyond the scope of this paper; however,
it is interesting to ask whether the spatial profiles of \lya\ emission from the same star-forming
galaxies can be understood in the context of a schematic model. 
In this section, we describe such a model that begins with inferences on the structure and
kinematics of CGM gas from S2010, and then test for consistency with both the \lya\ emission observations
and the absorption-based S2010 CGM model.

\subsection{A Model for \lya\ Scattering Halos}

We first consider the probability that a \lya\ photon produced in a galaxy's central few kpc 
will escape in the direction of a particular observer's line of sight. 
The escape probability will depend on the kinematics and optical depth distribution of 
the CGM gas, and so one might expect it to be closely related to  
the characteristics of absorption lines observable both in the galaxy spectra 
themselves ($b\simlt2$-3 kpc) and in lines of sight to background objects with 
at larger impact parameter  $b>>0$. 
The conditions necessary 
for a \lya\ line photon to escape in the direction of a particular observer 
are: 1) it must either be emitted at a frequency that is well
off resonance for any \ion{H}{1} in the foreground (i.e., between the point of emission and the
observer), and/or 2) it must be scattered in a direction that happens to have  
low spatial covering fraction $f_c$ of
\ion{H}{1}\footnote{In the limit of no \ion{H}{1} gas outside of a galaxy's \ion{H}{2} regions, the emergent \lya\ line would have
roughly the same spatial extent as that of the UV continuum.}.

\begin{figure}[thb]
\centerline{\includegraphics[width=9cm]{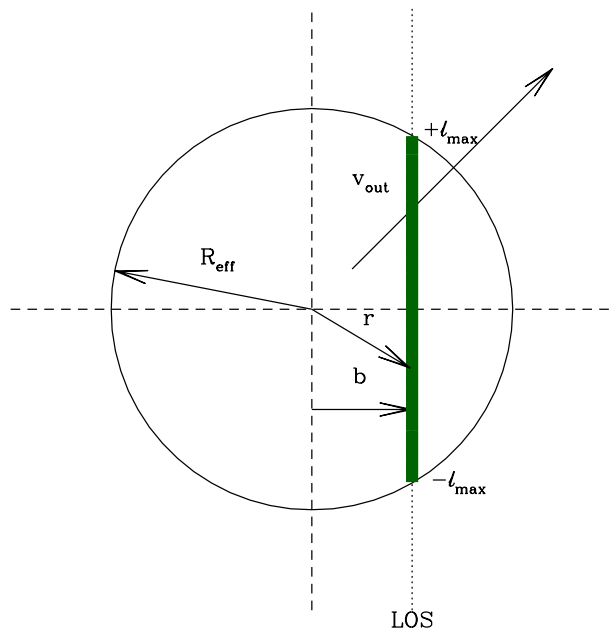}}
\caption{Coordinate system for the schematic model of scattered \lya\ emission from 
galaxies, where $b$ is the line of sight impact parameter, $r$ is the galactocentric distance, 
and $R_{eff}$ is the assumed physical size of the scattering medium. In the model, \lya\ photons are 
produced near $r \sim 0$, after which they diffuse outward until they are either destroyed or they
escape the scattering medium. 
The covering fraction of gas at galactocentric radius $r$ is assumed to be of the form $f_c(r) \propto r^{-\gamma}$ (see also S2010). 
\label{fig:coords}
} 
\end{figure}

For extended \lya\ produced by scattering in a gaseous halo, the observed surface brightness profile 
$S(b)$ will then be related to the integral along the line of sight at impact parameter $b$ of the product of a)
the \lya\ photon
density, b) the probability that a \lya\ photon will be scattered in our (the observer's) direction, and 
c) the probability that once scattered in our direction a photon will proceed to escape the nebula 
before being scattered once again. 
The situation is somewhat analogous to the galaxy outflow model used to match absorption 
line equivalent widths $W_0$ vs. impact parameter $b$
presented in S2010. Figure~\ref{fig:coords} shows the assumed geometry (cf. Figure 23 of S2010.) 
In the absorption case, a line of sight to a background object pierces the radial flow at
projected distance $b$, and the resulting absorption line strength is modulated by the integral along the line of sight
of the quantity $1-f_c(r,{\rm v_{out}})$, where $r$ is the galactocentric radius and ${\rm v_{out}}(r)$ 
is the flow velocity at radius $r$. 
As discussed by S2010, the velocity field in the absorbing gas can have a large effect on the strength of
absorption lines in the spectra of background sources when the transition is saturated, 
even if the covering fraction is significantly smaller
than unity. S2010 argued that consistency between the absorption line strength as a function of impact 
parameter on one hand, and the strength and profile shape of lines observed in the spectra of the galaxies themselves 
on the other,
requires large velocities and velocity gradients in the gas.  The absorption cross-section is dominated
by {\it outflowing} material, and  the flows 
are inferred to be clumpy (i.e., multi-phase), with  both high- and low-ionization ionic species observed over similar
ranges of velocity and galactocentric distance. In the context of the CGM model, most of the 
acceleration of cool gas to high velocity occurs in the 
inner several kpc. The covering fraction $f_c$ of gas giving rise to absorption in a particular transition decreases 
with increasing galactocentric distance $r$, modeled as a power law of the form $f_c \propto r^{-\gamma}$.

In general, the larger the range of gas-phase bulk velocity sampled along the observer's line of sight
at impact parameter $b$, the greater the chance
that a scattered \lya\ photon will reach the observer without further scattering. For simplicity, in
our model we assume that all scattering events are isotropic, and that the gas-phase velocity field is
axisymmetric and is a monotonic function of galactocentric distance $r$ (see S2010 for a justification of this assumption). 
If the bulk velocity field in the outflow has a range and amplitude much larger than that of the local velocity dispersion 
in the \ion{H}{1} gas\footnote{In the models discussed here, this is assumed to be the case based on
the results presented in S2010. While velocity is not used explicitly as a model parameter, large velocity
gradients along the line of sight directly affect the probability that a scattered \lya\ photon will
ultimately escape. In other words, the effective covering fraction of optically thick \ion{H}{1} as seen
by a \lya\ photon emitted from a particular position in the CGM  implicitly includes an integral over 
velocity even if it is not explicitly used as a model parameter. See \citealt{steidel2010} for a more
detailed discussion of this issue.}, or when 
when $f_c << 1$, the problem
can be reduced to a geometric one in which the covering fraction depends only
on galactocentric radius $r$, $f_c = f_c(r)$. Clearly it would be interesting to measure the velocity field
of extended \lya\ emission in order to gauge the role kinematics play in the transfer of \lya\
photons. Unfortunately, beyond the central, high surface brightness regions there are few constraints on the line
shapes, and at present we have only (projected) spatial information integrated over the full range
of velocity. 

When considering emission (rather than absorption in the spectra of background objects) 
one needs to account for the \lya\ ``source function'' which
varies with spatial position, as well as variations in opacity parametrized by $f_c(r)$. 
The \lya\ photon density available to contribute to the observed $S_{\lya}(b)$ will depend on
the fraction of \lya\ photons that have been able to diffuse outward to $r \simgt b$, which may be only a small
fraction of the \lya\ photons initially produced by recombination in \ion{H}{2} regions. 
When the covering fraction is high at small radii, 
one would expect the emergent \lya\ emission from that region to be suppressed -- photons are either
destroyed or radiatively trapped until they make their way to locations from which escape is more probable. 
The flux of \lya\ photons (assumed to be produced at small $r$ at a $\sim$ constant rate related to the SFR) 
at galactocentric radius $r$   
will be reduced by an overall geometric factor $1/4\pi r^{2}$, and by the destruction of \lya\ via 
absorption by dust grains\footnote{Under the assumption of spherical symmetry, \lya\ photons scattered at smaller
radii are returned to the ``pool'' of \lya\ photons potentially available for scattering at larger radii.} .

This diminished \lya\ radiation field would produce no \lya\ halo at $b > r$ if 
$f_c(r) = 0$, since photons would appear to be released from a \lya\ ``photosphere'' that would
approximately extend to the edge of the gas distribution. 
The apparent outer edge of the \lya\ scattering halo should correspond to the radius at which $f_c$ becomes
negligible and the scattered component of \lya\ falls below the observational threshold. 
At small radii, where $f_c(r)=1$ and the optical depth encountered in any direction is substantial,  
(e.g., for galaxies having absorption-dominated continuum spectra), \lya\ photons will be resonantly trapped
for a large number of scattering events before diffusing spatially outward. Most of the dust absorption, if present, 
would be expected to occur in such regions. Once $f_c(r)$ falls below unity at larger radii, \lya\ photons which have
not been destroyed may be scattered in the observer's direction. 
Thus, the relative rate of \lya\ scattering events at radius $r$ will be  
$ \propto f_c(r)/4\pi r^2$ where $f_c(r)$ is the \ion{H}{1} covering fraction. 
The chance that a scattered \lya\
photon will be emitted in the observer's direction (without any further interactions prior to escape) 
increases with decreasing 
characteristic $f_c$,  with probability roughly $\propto [1-f_c(r)]$ for $f_c(r) \le 1$. 
The \lya\ surface brightness as seen by an observer in a particular direction will
then be proportional to the product of these two terms, integrated along the line of sight through the galaxy
at impact parameter $b$:
\begin{eqnarray}
S_{\lya}(b) \propto S_0 \int_{-l_{max}}^{+l_{max}} \frac{f_c(r)[1-f_c(r)]}{4\pi r^2} dl ~~
\label{eq:sb_profile}
\end{eqnarray}  
where $l$ is the coordinate distance along the observer's line of sight at impact parameter $b$, 
$l_{max}=(R_{\rm eff}^2-b^2)^{\frac{1}{2}}$, $S_0$ is a normalization for the surface brightness distribution,  
and $R_{\rm eff}$ is the effective size of the scattering halo (see Figure~\ref{fig:coords}) .  
Note that the integrand tends toward zero when $f_c(r) \simeq 1$, qualitatively 
accounting for the suppression of \lya\ emission in regions with high $f_c(r)$.  
Clearly, when $f_c=1$ the purely geometric model is no longer valid, since 
the \lya\ emission intensity associated with an optically thick region can be ``negative'', 
i.e., there is a net removal of \lya\ photons at that spatial position that will reduce the
net surface brightness along that particular line of sight. 

In the S2010 CGM model, 
the radial dependence of the covering fraction of gas was found to be consistent with 
a power law of the form $f_c(r) \propto r^{-\gamma}$, where $0.3 \simlt \gamma \simlt 0.6$ depending on the ionization state of the
tracer ion; $f_c(r)$ becomes consistent with zero for $r > R_{\rm eff} \sim 90$ kpc for most of the observed ions. 
If we take the power law form for $f_c$ (with index $\gamma$) 
characteristic of the highest \ion{H}{1} optical depth material, 
equation~\ref{eq:sb_profile} can be used to predict $S_{\lya}(b)$ given an overall normalization $S_0$, 
a characteristic radius $r_0$ at which $f_c(r)$ first falls 
below unity [i.e. $f_c(r) =(r/r_0)^{-\gamma}$], and the the effective size of the scattering medium, $R_{\rm eff}$.  
We account qualitatively for the variation of the spatial profile of \lya\ emission 
in the central regions of a galaxy 
by (artificially) allowing $f_c > 1$ at $r < r_0$, using an extrapolation of the same power law form for
$f_c(r)$. Since the integrand becomes negative when $f_c > 1$, this leads to suppression
of the \lya\ surface brightness for any line of sight that intercepts such a region. 
In practice, the central \lya\ emission must be substantially suppressed to match the observed profiles of any
of the galaxy subsets, including the LAEs 
(cf. Figure~\ref{fig:sb_profile}).

\begin{figure}[thb]
\centerline{\includegraphics[width=9cm]{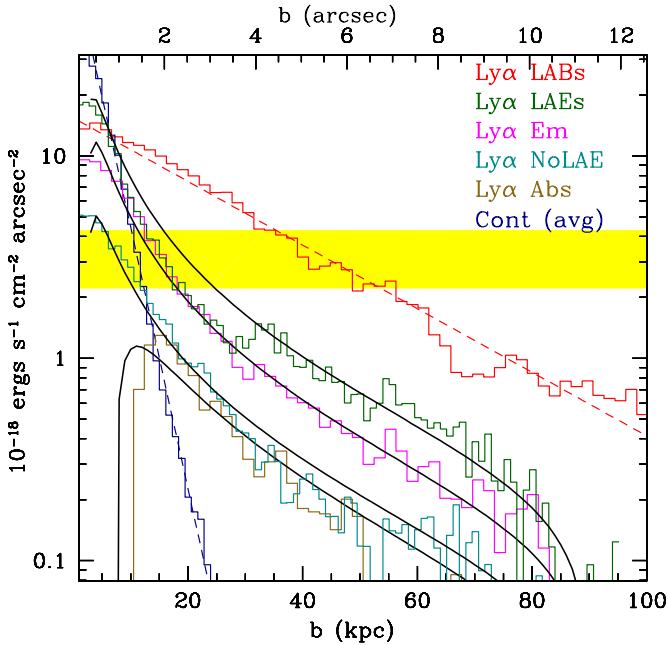}}
\caption{Same as Figure~\ref{fig:blob_plot}, but where models described by equation~\ref{eq:sb_profile} have
replaced the exponential profile used in Figure~\ref{fig:blob_plot}.  
Parameters for the 4 models shown are summarized in Table~\ref{tab:models}. 
All 4 model curves have $R_{\rm eff} = 90$ kpc and 
$f_c(r) \propto (r/r_0)^{-\gamma}$ with $\gamma \simeq 0.6-0.8$. The ``\lya\ Abs'' model  
produces a central hole in the \lya\ emission by adjusting the normalization of $f_c$ 
with the parameter $r_0$; the larger value of $r_0$ indicates that the CGM remains
optically thick to \lya\ photons to larger galactocentric radii than for the other
sub-samples.  
\label{fig:blob_plot_mod}
}
\end{figure}

\begin{figure}[thb]
\centerline{\includegraphics[width=9cm]{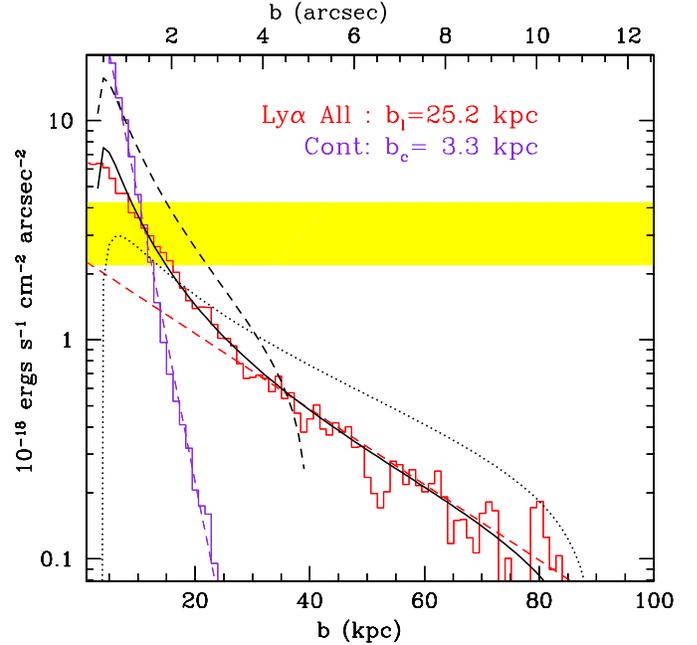}}
\caption{Same as Figure~\ref{fig:sb_profile}, where the preferred model is drawn with the
solid black curve, corresponding to $S_0 = 11.5\times10^{-18}$ ergs s$^{-1}$ cm$^{-2}$ arcsec$^{-2}$, $R_{\rm eff} = 90$ kpc, $r_0 = 2.2$ kpc,
and $\gamma = 0.6$,
for covering fraction parametrized as 
$f_c(r) \propto (r/r_0)^{-\gamma}$ (see also S2010). The dashed curve shows a model 
with the same parameters, except 
$R_{\rm eff} = 50$ kpc, while the dotted curve assumes $\gamma = 0.2$, with all other parameters as for the preferred 
model. 
\label{fig:plot_all_demo}
}
\end{figure}

Figures~\ref{fig:blob_plot_mod} and~\ref{fig:plot_all_demo} show example models based on equation~\ref{eq:sb_profile}
compared with the observed composite
\lya\ surface brightness profiles; the corresponding model parameters are given in 
Table~\ref{tab:models}. 
The overall shape of the predicted SB profile is sensitive to the value of $\gamma$ parametrizing the radial dependence
of the covering fraction of \ion{H}{1}; $\gamma < 0.3$ produces \lya\ profiles that are 
flatter than observed, while
$\gamma > 0.8$ predicts \lya\ emission which falls too rapidly with increasing $b$ (Figure~\ref{fig:plot_all_demo}). 
As discussed above, the shape of the central portion of $S_{\lya}(b)$ is modulated by adjusting the galactocentric
radius $r_0$ where $f_c(r_0) =1$ (i.e., $r_0$ serves as a normalization of the maximum covering fraction).
The presence of a central ``hole'' in $S_{\lya}(b)$ (as observed for the \lya\ Abs sub-sample in
Figure~\ref{fig:blob_plot_mod}) 
can be reproduced by increasing $r_0$ so that the transition from net \lya\ absorption to net \lya\ 
emission moves to larger galactocentric radius.  Once \lya\ has a finite probability of escape 
(i.e., where $f_c(r) < 1$ in the context of our simple model), the residual \lya\ photons at $r > r_0$ become available for
re-direction toward an external observer who then perceives the photon being ``emitted'' from a position at impact
parameter $b$ in projection.

The values of $\gamma$ required to produce model profiles in reasonable agreement with their observed counterparts 
(Table~\ref{tab:models} and Figure~\ref{fig:blob_plot_mod}) are near the high end of the range inferred from 
the behavior of absorption line strength $W_0$ versus impact parameter $b$ (S2010). One possible explanation for
slightly steeper profiles is that the emission models assume {\it no} \lya\ photons are destroyed once they
propagate beyond $r \simeq r_0$; if \lya\ has a finite chance of being absorbed by dust at $r > r_0$, the additional
attenuation of \lya\ would manifest itself as a steepening of the profile with respect to the pure scattering
model.  That the \lya\ Abs model exhibits both the steepest decline in
\lya\ surface brightness ($\gamma \simeq 0.8$ compared to $\gamma \simeq 0.6$ for the other sub-samples) 
and the largest global extinction correction (\S3 and Table~\ref{table:tab3}) suggests dust may not be confined 
solely to the central regions in such galaxies. 

The model of the CGM proposed by S2010 almost certainly does not provide a unique explanation for the
IS absorption line strength and kinematics as observed in the spectra of background galaxies; 
however, we have shown, with a simple extension of the  
model, that scattering of \lya\ photons from the same CGM gas can  
can also account for \lya\ emission with radial surface brightness profiles and physical extent consistent with the observations. 
Regardless of the model details (which admittedly could be incorrect), 
the very similar physical scales involved ($R_{\rm eff} \simeq 90$ kpc) suggest 
a close causal connection between the cool gas observed to produce strong \ion{H}{1} and low-ionization metallic 
absorption lines in the spectra of background continuum sources, {\it and} spatially extended \lya\ emission from the
same host galaxies.  
\begin{deluxetable}{l c c c c }
\tablewidth{0pt}
\tabletypesize{\scriptsize}
\tablecaption{Parameters for Model \lya\ Spatial Profiles}
\tablehead{
\colhead{Sample\tablenotemark{a}} &
\colhead{$S_0$\tablenotemark{b}} &
\colhead{$r_0$ (kpc)\tablenotemark{c}} &
\colhead{$\gamma$\tablenotemark{d}} &
\colhead{$R_{\rm eff}$ (kpc)\tablenotemark{e}} 
} 
\startdata
All & 11.5 & 2.2 & 0.6 & 90  \\
Ly$\alpha$ Em & 17.0 & 2.0 & 0.6 & 90 \\
Ly$\alpha$ Abs & ~4.5 & 5.9 & 0.8 & 90 \\
All non-LAE & ~7.0 & 2.0 & 0.6 & 90 \\
LAE only & 25.0 & 1.9 & 0.6 & 90 \\
\enddata
\tablenotetext{a}{Galaxy sub-sample, drawn from the full sample (All) of 92 continuum-selected
galaxies with Ly$\alpha$ imaging. }
\tablenotetext{b}{Intensity normalization for model (see Eq. \ref{eq:sb_profile}), in units of $10^{-18}$
ergs s$^{-1}$ cm$^{-2}$ Hz$^{-1}$.} 
\tablenotetext{c}{Galactocentric radius at which $f_c = 1$}
\tablenotetext{d}{Power law index in the radial behavior of the covering fraction, $f_c = (r/r_0)^{-\gamma}$.}
\tablenotetext{e}{Effective size of CGM region producing detectable \lya\ emission, in kpc.}
\label{tab:models}
\end{deluxetable}

\subsection{Comparison with \lya\ Emission in Simulations}

A scenario in which extended \lya\ emission around galaxies is {\it dominated} by scattering of \lya\ photons
initially produced inside the galaxies, rather than by external processes, has 
been the focus of a number of recent galaxy models including treatment of \lya\ radiative 
transfer (e.g., \citealt{verhamme08,laursen09a,laursen09b,zheng10a,zheng10b,barnes11}). 
Each of these studies places emphasis on different aspects of the model galaxies, and 
cursory examination suggests a qualitative
similarity to the observations presented here, since \lya\ scattering leads to the spatial
re-distribution of the \lya\ emission as seen by an observer.  
Gas-phase kinematics play a large role in determining how much \lya\ emission escapes the
galaxies, and in all of the models except those of \citealt{verhamme08} (which 
do not explicitly consider the spatial distribution of \lya\ emission) the dominant velocity field is
associated with infall/accretion. The predicted \lya\ line profiles tend to be asymmetric 
and sometimes double-peaked, usually dominated by photons that are blue-shifted with respect to the
galaxy systemic velocity-- a configuration that is very rarely observed in galaxy spectra (e.g., \citealt{pettini00,
shapley03,steidel2010}). Also, while the 3-D models all produce \lya\ emission
that is significantly more extended than the UV continuum, the predicted surface brightness
profiles of scattered emission generally declines much more rapidly than for the
observed LBG \lya\ halos (i.e., most would fall well below the current surface brightness limit). 
\cite{barnes11} have pointed out that higher outflow velocities tend to produce more extended
\lya\ emission in the context of their models, which include both inflows and outflows of gas, 
so that perhaps the missing ingredient is the presence of higher-velocity outflows than have
generally been modeled. 

Most of the simulations work on \lya\ emission from galaxies has not highlighted the 
potential utility of using \lya\ emission observations as a means of revealing gas-phase 
structure in the surrounding CGM and IGM.
One exception is a
a series of recent papers exploring
how very sensitive NB observations of \lya\ emission can be used along with cosmological simulations (including detailed
radiative transfer) to trace the underlying 
large scale structure at high redshifts (\citealt{zheng10a,zheng10b}). In \cite{zheng10b}, the authors explicitly
calculate the expected properties of diffuse \lya\ emission around star-forming galaxies, with principal focus
on LAEs at $z \simeq 6$.  Like the scenario we have described above, the models assume that the 
ultimate source of the \lya\ photons seen in emission 
is the galaxy HII regions, with extended emission resulting from the details of the \lya\ radiative transfer. 
\cite{zheng10b} predict that the surface brightness profile surrounding individual galaxies will have 
two distinct components related closely to 1) the ``halo exclusion scale'' within comoving distances of $0.3h^{-1}$ Mpc 
($b \simeq 60$ physical kpc at $z \simeq 6$), and a larger-scale component arising from galaxy clustering, extending to $\simeq 3$ Mpc (comoving),
or $\simeq 400$ physical kpc at $z \simeq 6$. 
The smaller scale is similar to the virial radius of the characteristic dark matter halos   
being considered in the simulation.
\footnote{Note that our procedure
of masking out all identified continuum sources other than the central one when producing the \lya\ and continuum
stacks (\S3) would suppress what \cite{zheng10b} call the ``two-halo term'' due to clustering, so our observed
\lya\ profiles should be compared only with the ``one-halo'', central component. }
It is not completely straightforward to move the predictions to
$ z \simeq 2.65$ for comparison with our observations, but (as discussed in S2010) the CGM scattering medium 
observed around $z \sim 2-3$ galaxies (which we have argued is responsible for absorption against background sources
as well as for the extent of scattered \lya\ emission) also has a size similar to the virial radius $r_v\simeq 80-100$ kpc.
However, it is not clear that the observations are consistent with the predictions when it comes to the dependence of
the \lya\ emission halo on other properties of the galaxies. In the \cite{zheng10b} models, the primary driver of
the surface brightness profile is the gas-phase kinematics of the CGM gas; the characteristic
scale of the inner component of \lya\ emission relates to the ``infall'' region for the halo, within which 
gas is accreting onto the central galaxy (the simulations do
not have outflowing material, and it is the kinematics of infalling material
that modulate the escape of \lya\ photons). On the other hand, in our picture the 
characteristic scale is related to the radial dependence of the 
covering fraction of neutral material {\it and}
the gas-phase kinematics (assumed to be dominated by outflows). Within our sample of LBGs, the stacked \lya\ images
of various subsets indicate a rather consistent exponential scale length of $b_l \simeq 25$ kpc with at most
a weak dependence on \lya\ or UV luminosity or on the fraction of \lya\ photons that escape the galaxy ISM. 
If there is a trend, it is in the direction opposite to that expected in the models. 

Interestingly, many recent theoretical investigations focusing primarily on diffuse and
extended \lya\ from the outer parts of galaxies or LAEs  
have deliberately neglected the scattering of \lya\ from the inside out (\citealt{dijkstra06a,dijkstra09,keres08,fauch10,goerdt10}). 
Instead, attention has been drawn to \lya\ emission associated with gas cooling as it accretes onto galaxies 
(``\lya\ Cooling''), 
or on \lya\ fluorescence as a means of measuring the intensity
of sources of ionizing photons at high redshifts.  Both of these processes are discussed in \S5 below.  

In any case, there is no doubt that radiative transfer
calculations will be key to a full understanding of diffuse \lya\ emission from galaxies. However, it is essential that
the CGM gas distribution and kinematics in the simulations match real galaxies. Without the correct gas-phase
model, even the most sophisticated treatment of radiative transfer cannot yield a realistic result. The observations
suggest that possibly important ingredients include a CGM that is clumpy on small scales and which has very
large (non-gravitational) velocity gradients dominated by galaxy-scale outflows.  

\section{Discussion}

We have shown above that, on average, LBGs with far-UV luminosities $0.3\simlt (L/L_{UV}^*) \simlt 3$ at 
$\langle z \rangle = 2.65$ exhibit spatially extended \lya\ emission
to physical radii of at least 80 kpc (10\arcs), even when \lya\ appears only in absorption for regions coincident with
the UV continuum starlight.  Figures~\ref{fig:sb_profile}, \ref{fig:emabs_plot}, and \ref{fig:blob_plot_mod} 
show that the {\it profiles} of the \lya\ emission are quite similar in shape independent of the spectral morphology,
with the main difference being the overall intensity normalization and the presence or absence of emission spatially coincident
with the continuum light (i.e., the inner $\pm 5$ kpc). The observations suggest that the \lya-scattering CGM 
may be statistically universal, with the main variable being the fraction of \lya\ photons able to emerge from 
the inner few kpc region without being destroyed. For example, the difference between the \lya\ Em and \lya\ Abs
(see Table~\ref{table:tab2}) spectrally classified subsets is an overall factor of $\sim 5$ in the \lya\
surface brightness at the full continuum extent (Figure~\ref{fig:emabs_plot}), beyond which the ratio 
of $S(b)$ for the two sub-samples remains essentially constant. 
The scale lengths for \lya\ emission ($b_l \simeq 25\pm3$ kpc) are consistent among the statistically distinct galaxy sub-samples 
in spite of the fact that the integrated line-to-continuum ratio varies by large factors among the same sub-samples. 

\subsection{Previous Results on Statistical \lya\ Detections}

\begin{figure}[thb]
\centerline{\includegraphics[width=9cm]{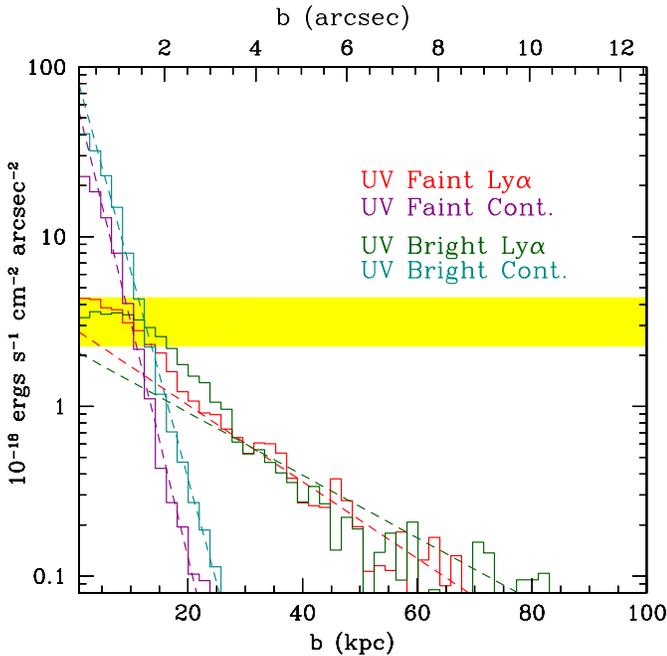}}
\caption{A comparison of the continuum and \lya\ surface brightness profiles of the full sample
divided into two at the median continuum apparent magnitude. The ``UV bright'' sample
is a factor of $\simeq 2.0$ times brighter in the continuum than that of the ``UV faint'' sample
(CB(Bright)$=24.22$ versus CB(Faint)$=24.95$), but the average \lya\ flux for the UV bright
sub-sample is 10\% {\it smaller} than that of the UV-faint sub-sample, i.e. $W_0(\lya,bright)=22.0$ \AA,
while $W_0(\lya,faint)=48.5$ \AA).  
\label{fig:Rs_compare}
}
\end{figure}

\lya\ emission with physical extent larger than that of a galaxy's continuum starlight is not 
a surprising result from a theoretical perspective (e.g., \citealt{barnes09,barnes10,laursen09a,laursen09b}), and has been observed and
noted in many individual cases both in the nearby (e.g. \citealt{mas-hesse03,hayes07,ostlin09})
and high redshift (e.g.,\citealt{franx97,moller98,steidel00,fynbo03,matsuda04,adelberger06,ouchi08}) universe. 
However, relatively few surveys at high redshift have
reached adequate \lya\ surface brightness limits to allow the detection of the very low surface brightness levels    
discussed above.  An exception is the extremely deep spectroscopic survey for \lya\ emission conducted by
\cite{rauch08} [R08]. Using a \lya\ --selected sample distributed over the redshift range $2.7 \le z \le 3.8$, 
these authors noted that extended \lya\ emission was a common feature of the LAEs discovered in their
survey. A spatial stack of all of the \lya\ emitting sources exhibited significant emission (with threshold $\simeq 1.5\times
10^{-19}$ ergs s$^{-1}$ cm$^{-2}$ arcsec$^{-2}$) to an angular scale of 
$\sim 4$\arcs, or $\sim 30$ kpc projected physical radius. The R08 sample, as the authors themselves point out,
covers a different range of UV luminosity compared to most continuum-selected LBG spectroscopic surveys-- only one of 27 objects 
has $V < 25.5$, while 80\% our sample (which has a median $V\simeq 25.0$) has $V < 25.5$, although there is
is a tendency for the faintest objects to be among those with the strongest \lya\ emission lines 
(see Tables~\ref{table:tab2} and ~\ref{table:tab3}). 
\footnote{Moving our continuum-selected sample to the somewhat higher 
median redshift of R08 would result in $\simeq 50$\% of our sample having $V > 25.5$. }
Nevertheless, the average surface brightness profile for the R08 \lya-selected sample is remarkably
similar to that of our continuum-selected sample (e.g., compare Figure~\ref{fig:emabs_plot} to Figure 20 of R08). 
For objects in our 
``\lya\ Em'' sub-sample (Table~\ref{table:tab2}), the peak \lya\  
SBs are somewhat higher than for the R08 sample, while the angular extent (at the same limiting SB of $\sim 1\times10^{-19}$
ergs s$^{-1}$ cm$^{-2}$ arcsec$^{-2}$) is $\simeq 2.5-3$ times larger in the present LBG sample. 
Within our sample there is a significant dependence of $W_0(\lya)$ on apparent UV continuum luminosity,
but the average \lya\ profiles are similar, as shown in Figure~\ref{fig:Rs_compare}.

In any case, it is worth
pointing out that, under the hypothesis that \lya\ scattering, and not fluorescence, is the dominant process
producing the observed \lya\ halos, the scattering medium need not be optically thick in the \ion{H}{1} Lyman continuum.
This means that it is not necessarily correct to associate the observed physical extent of \lya\ emission with regions 
having ${\rm N(HI) > 3 \times 10^{18}}$ cm$^{-2}$ as R08 have suggested -- in principle, ${\rm N(HI)}$ could be 1000 times lower and still remain
optically thick to \lya\ photons. 

Perhaps more directly analogous to the results of the present sample is the narrow-band \lya\ survey of \cite{hayashino04}. These
authors used deep NB \lya\ images in  
the SSA22 field, and stacked the \lya\ images of 22 $z =3.09$ continuum-selected 
LBGs from the survey of \cite{steidel03}, of which 19 are in common with our current SSA22 sample\footnote{   
The new NB image used in the present sample includes both archival Subaru data as well as an additional 10 hours'
integration using LRIS on the Keck 1 telescope, and so is substantially
deeper ($\sim$ factor of 2-3), but covers a much smaller area, than that of \cite{hayashino04}. }.    
Indeed, Hayashino et al showed that significant emission extends to angular scales of at least
4\arcs\, and that the ``ring'' in the range 2-4\arcs\ often contains as much or more \lya\
flux than the inner $\theta \le 2$ \arcs\ region. They also stated (but did not show) that a stack of the 13 galaxies
which did not individually exhibit extended \lya\ emission results in a significant detection on the same
2-4\arcs\ scales. Although the authors did not discuss what physical mechanism might have been responsible
for their observation, these results clearly provided an early indication of the nature of \lya\ emission in 
L* galaxies, borne out by our larger and more sensitive sample.

\subsection{Has the Whole Iceberg Been Detected?}
   
The level of sensitivity to low-SB \lya\ emission at high redshifts is unlikely to improve by large
factors using the current generation of ground-based telescopes, and so a natural question would be:
How much more is there at still lower SB? 
Many \lya\ surveys (e.g., \citealt{rauch08,bunker98}) have been designed to 
detect \lya\ fluorescence induced by the metagalactic radiation field at redshifts $2 \simlt z \simlt 3$.
The radiation field intensity is 
usually expressed as $J_{\nu} \simeq 2-10 \times 10^{-22}$ ergs s$^{-1}$ cm$^{-2}$ Hz$^{-1}$ sr$^{-1}$, where
the quoted range indicates the dispersion among published observational or theoretical estimates (e.g., \citealt{shapley06,
bolton05,scott00,fauch08}). 
The expected maximum fluorescent signal at $z \simeq 2.5-3.0$ is in the range $0.2-2 \times 10^{-19}$ ergs s$^{-1}$
cm$^{-2}$ arcsec$^{-2}$ if the only source of ionizing photons is
the general UV background (see e.g. \citealt{cantalupo05,kollmeier10,fauch10}). 
These expectations clearly lie at or below the current SB thresholds of any survey
completed to date. The difficulty of detecting the fluorescent signal from the metagalactic UV field has instead inspired several
searches for fluorescence near bright sources of ionizing photons, such as QSOs (\citealt{francis04,cantalupo05,
adelberger06,hennawi09}). The results from such studies have been mixed. 

A different argument can be used to suggest that fluorescence from the UV background will always be overwhelmed
by \lya\ scattering from the CGM of star-forming galaxies, at least at $z \sim 2-3$. This assertion follows 
from the fact that S2010 found that the total absorption cross-section contributed by the CGM of LBGs (using
${\rm R_{eff}} = 80-90$ kpc for the detection of low-ionization absorption species) can account for a large fraction of 
all gas with  
N(\ion{H}{1})$\simgt 2\times10^{17}$ cm$^{-2}$ (i.e., $\tau \ge 1$ in the Lyman continuum, also known as
``Lyman Limit Systems''). In other words, any gas of sufficiently high $N(HI)$ to produce a detectable signal 
from fluorescence also lies within $\sim 90$ kpc of a star-forming galaxy with properties similar to those
in our sample.   
We have shown that these galaxies generically exhibit diffuse \lya\ emission on the same physical scales when 
a surface brightness threshold of $S(\lya) \sim 1 \times 10^{-19}$
ergs s$^{-1}$ cm$^{-2}$ arcsec$^{-2}$ is reached. Unless the fluorescent \lya\ signal lies at the very
top of the allowed range, it will have much lower SB than the signal we have attributed to scattering from the inside
of the galaxy out. 

It is more difficult to assess what fraction of observed \lya\ emission may be due to cooling 
processes such as those described by a number of recent authors (e.g., \citealt{dijkstra09,kollmeier10,fauch10,goerdt10}.) 
In particular,
the predictions of the emergent \lya\ emission from cooling gas accreting onto galaxies are extremely sensitive to 
gas temperature (\citealt{kollmeier10,fauch10}) and to the small-scale structure in the gas. As a result,
the range in \lya\ flux and SB, as well as the galaxy mass dependence and spatial distribution of cooling emission,
must be regarded as uncertain by a factor of $\simgt 10$, with an upper bound (based on energetic arguments)
that can be as large as $L(\lya) \simgt 10^{44}$ ergs s$^{-1}$, but which under different assumptions could
be as small as $\sim 5 \times 10^{41}$ ergs s$^{-1}$ for a galaxy with $M_{halo} \simeq 9 \times 10^{11}$ 
M$_{\sun}$ (\citealt{fauch10}), approximately the mean halo mass of the galaxies in the present sample (see \citealt{asp+05,conroy08,steidel2010}). 

The observations appear to argue against a significant contribution of cooling radiation to the detected
\lya\ halos, at least on average. We have shown that the shape of the observed radial surface brightness distribution  
among the LBGs in the sample is remarkably consistent beyond the inner $\sim 10$ kpc, within which the
\lya\ intensity for a given continuum luminosity varies by orders of magnitude. Moreover, the overall intensity
scaling for the \lya\ emission at large radii is strongly correlated with the behavior of \lya\ emission in
the inner 5-10 kpc region --- at the same continuum luminosity, \lya\ absorption-dominated galaxies (on average) exhibit
diffuse \lya\ emission with a factor of
3-4 lower normalization than galaxies with spectroscopically detected \lya\ emission.  
In the context of \lya\ cooling radiation, one might expect the extended \lya\ emission to be strongly
correlated with galaxy mass and/or SFR since it is believed by some (e.g., \citealt{goerdt10})
that the baryonic accretion rate ultimately controls the SFR. 
In this scenario, the central region of \lya\ emission might be suppressed by higher \ion{H}{1} column densities
mixed with dust, but the outer regions would have no obvious way to ``know about'' the number of \lya\ photons
being produced at smaller radii. Instead, 
one might expect that the brightest \lya\ halos would be associated with the ``\lya\ Abs'' sub-sample, since these have a median
SFR nearly 3 (4.5) times larger than the ``\lya\ Em'' (LAE) sub-samples. 
Clearly, the observations are inconsistent with this expectation. 
If on the other hand most or all of the \lya\ emission at all radii originates 
in the central regions and is subsequently scattered by the CGM gas, the density of photons available for scattering at (for example)
$r=50$ kpc will be very tightly linked to the number of \lya\ photons that successfully diffuse past
$r \sim 5$ kpc, beyond which the \lya\ halos appear ``self-similar''. 
The emergent \lya\ luminosities are entirely consistent with the observed 
level of star formation in the galaxies, and are more attenuated than the UV
continuum, for all sub-samples except the LAEs. It
is not necessary to invoke sources of \lya\ emission other than scattering (from the inside outward) 
to account for both the \lya\ luminosity and its spatial distribution. 

Under the scattering hypothesis, and further assuming that the scattering medium is self-similar for all galaxies, 
then sub-samples with more luminous \lya\ halos should provide information on the degree to which even the current SB
threshold might lead to an underestimate of the total \lya\ flux emergent from a galaxy. To increase the dynamic
range for detecting diffuse \lya\ emission, one might use the observed properties of giant LABs (which are 
well-detected in the stack to $b\simeq 15$\arcs) to estimate how much additional \lya\ flux may lie 
beyond the SB detection threshold near $b \sim 8$\arcs\ for more typical galaxies. Under the assumption that 
diffuse emission from LABs and LBGs has a similar origin and differs only in total \lya\ luminosity, the 
curve-of-growth for LABs (Figure~\ref{fig:lya_cum_plot}) suggests that an aperture of radius $\simeq 8$ \arcs\
would underestimate the total \lya\ flux by only $\sim 10$\%.  
Thus, further aperture corrections to the integrated \lya\ would probably leave the values of $W_0(\lya)$ (Table~\ref{table:tab2})
and $f_{esc,rel}(\lya)$ 
(Table~\ref{table:tab3}) more or less unchanged.  At least at $z\simeq 2.65$, the current SB limit
appears to be sufficient to detect most of the ``iceberg''. 
    
Finally, we note that the differences in the intensity of the large-scale diffuse emission among sub-samples
divided according to their spectral morphology suggest that galaxy viewing angle is relatively unimportant (on average) 
for \lya\ emission; that is, most  
galaxies are not LAEs in some directions but strong \lya\ Abs systems in others, consistent with the inference
of generally axisymmetric CGM gas distributions inferred from the absorption line studies (S2010).    

\subsection{IS Absorption, \lya\ Emission, and the CGM}

Perhaps the strongest correlation (first explored in detail by \citealt{shapley03} for galaxies
at $z \sim 3$) among the observed spectral properties of LBGs is between
the strength of low-ionization IS absorption lines and 
the spectral morphology and equivalent width of \lya. 
Galaxies with the strongest
\lya\ emission (among the continuum-selected samples) invariably have much {\it weaker} than
average low-ionization IS absorption lines (see \citealt{erb2010} for a well-observed example),
while those with the \lya\ appearing strongly in absorption have correspondingly strong 
IS absorption features, often reaching zero intensity over some or most of the line
profile (see e.g. \citealt{prs+02}) indicating unity covering fraction. 
These trends are easy to understand in the context of the CGM model discussed by S2010 and
extended in this paper to cover the expectations for \lya\ scattering and its effects 
on the observability of \lya\ emission: both the IS absorption lines and \lya\ line strengths
and morphologies are controlled by the kinematics and geometry of the same interstellar and circum-galactic gas. 

Dust certainly plays a role in determining the fraction of both \lya\ and continuum photons 
that will end up reaching an observer. However, the gas-phase geometry and kinematics are more directly responsible
for the observed line strength (and line-to-continuum ratios) in the spectra. If a galaxy has strong \lya\ emission emerging from the same 
region as the UV continuum, it {\it must} have shallow IS absorption lines; if it did not, then at least the
spatial distribution of \lya\ (if not also its integrated flux) would be substantially modified--  it
would become more spatially diffuse. When a slit spectrum (generally a small-aperture measurement)
shows very strong and deep low-ionization
IS absorption lines, including \lya, it {\it must} be the case that any \lya\ seen in emission will have 
escaped either from a region spatially distinct from the continuum (the subject of this paper), 
or by way of scattering from very high velocity material (see S2010).  \lya\
emission seen in spectra which also show strong IS absorption will be primarily in the latter category, hence
the nearly universal systemic redshift of \lya\ emission in LBG spectra. We have emphasized above that
any \lya\ photons that are not destroyed by dust will eventually find their way out of their host galaxy-- 
but will be much harder to detect by the time they do.   

The point is that IS absorption and \lya\ emission are causally intertwined through their mutual dependence on the structure
and kinematics of the CGM on scales from a few kpc to $\simeq 100$ kpc.  

\section{Summary}

We have presented observations of a sample of 92 continuum-selected LBGs at 
$\langle z \rangle = 2.65$
having both rest-UV spectra and very deep  
narrow-band \lya\ images. The sample, which is representative of $\simeq L*$ LBGs at $2 \simlt z \simlt 3$,
was used to examine the nature of diffuse \lya\ emission from star-forming galaxies as function of 
their spectral morphologies and NB-inferred \lya\ fluxes.  
By stacking both UV continuum and \lya\ line images for subsets of the galaxy sample, we are able
to study the spatial distribution of \lya\ and continuum emission to much lower surface brightness
thresholds ($\sim 1 \times 10^{-19}$ ergs s$^{-1}$ cm$^{-2}$ arcsec$^{-2}$)
than would be possible for individual galaxies. We find:

1.  Relatively luminous star-forming galaxies generically exhibit low-surface brightness 
\lya\ emission to projected radii of at least 80 physical kpc ($\sim 10$ arcsec).
The extended emission is present even for the stacks of LBGs that would be classified
based on their spectra as having \lya\ in net absorption. 
The \lya\ line to UV continuum ratio is always strongly suppressed in the central regions of
galaxies relative to ``Case B'' expectations, but beyond galactocentric radii of $r \simeq 5$ kpc,
where the continuum light falls off very rapidly, \lya\ emission begins to dominate. 

2.  The \lya\ emitting regions have characteristic exponential scale lengths 5-10 times larger than the 
the corresponding UV continuum emission {\it from the same galaxies}.   
It appears that on average {\it all} classes of star-forming galaxies in the observed range of luminosity would
be classified as ``Lyman $\alpha$ Blobs'' if the observations were sufficiently sensitive. 
Similarly, nearly all galaxies would also be classified as ``\lya\ Emitters'' if their
total \lya\ flux were measured using a sufficiently large aperture.  
Spectroscopic measurements (or, typically deep \lya\ narrow-band surveys) of \lya\ 
emission underestimate the total \lya\ flux [or, equivalently, the rest equivalent
width $W_0(\lya)$] by an average factor of $5$, and a factor of $>3$ even for 
those classified as LAEs. 

3.  The surface brightness distribution, total flux, and scale lengths for \lya\
emission are all consistent with a picture in which most or all detectable \lya\ emission 
is produced in \ion{H}{2} regions spatially coincident with the galaxies' UV continuum emission. 
The \lya\ surface brightness is then modified by scattering from the surface of \ion{H}{1} 
clouds that are being driven to large galactocentric radii by galaxy-scale outflows. The spatial
extent of observed \lya\ emission is then dictated by the spatial extent of circum-galactic 
gas with sufficiently large covering fraction to have a finite chance of scattering \lya\
photons in the direction of an observer. 

4. The inferred attenuation of \lya\ emission from continuum-selected LBGs is consistently larger than that of
the UV continuum ($A(\lya) \simeq 1.6 A(UV)$) for all sub-samples except LAEs, which have $A(\lya) \simeq A(UV)$.
Most of the attenuation of \lya\ emission appears to occur within $\sim 5$ kpc
of the continuum centroid of a galaxy. While the fraction of a galaxies' total \lya\ photon production that
is able to diffuse beyond $\sim 5$ kpc varies substantially, the scattering halo of 
cool material
at larger radii leads to self-similar diffuse \lya\ halos. A simple scattering model for \lya\
emission was presented, based on the structure of the CGM gas inferred from 
measurements of absorption in lines of sight passing within $b  < 125$ kpc of
an ensemble of similar galaxies. The model successfully accounts for both 
the typical size and the surface brightness profile of \lya\ emission.  

5. We argue that scattering of \lya\ photons from circum-galactic gas can account for all
of the observations of continuum-selected star-forming galaxies, and that the observed
correlations of the intensity of diffuse \lya\ halos with the spectral morphology of the
central galaxies argues against a significant contribution from \lya\ cooling of accreting
gas.  We also argue that \lya\ scattering processes will always dominate
over fluorescence (caused either by the metagalactic ionizing radiation field or by ionizing 
photons from inside the galaxy) in producing spatially extended \lya\ emission.  

6.  \lya\ emission and interstellar absorption line strengths are causally intertwined
through their mutual dependence on the structure and kinematics of CGM gas. Galaxies
with strong and centrally-peaked \lya\
emission are expected to be associated with shallow IS absorption lines, while strong \lya\ absorption
lines that completely absorb the UV continuum light of the host galaxy force \lya\ emission to larger
galactocentric radii before escaping the host galaxy.  
 
The detection of diffuse \lya\ emission halos at the current surface brightness level has required 
the equivalent of $\simeq 200-1000$ hours' integration time with a 10m-class telescope (accounting for
the effective integration time of the stacked \lya\ line images). Given the
$(1+z)^{-4}$ dependence of observed \lya\ surface brightness (for a given physical luminosity surface
density), it is not feasible at present to obtain similar results at significantly higher
redshifts. However, vastly increasing the number of continuum-selected LBGs with sensitive NB \lya\
observations could in principle trace \lya\ emission from galaxy halos until they become
indistinguishable from the background. 
Together with observations of the cool gas phase via absorption lines in the spectra of background sources 
(both galaxies and QSOs),
such \lya\ emission observations considerably enhance our ability to observe directly the distribution of cool baryons
and their flow rate into and out of forming galaxies during an undoubtedly crucial (but not well-understood) period in cosmic history. 

\acknowledgments
 
This work has been supported by the US National Science Foundation through grants
AST-0606912 and AST-0908805 (CCS), and by the David and Lucile Packard Foundation (AES).
CCS acknowledges additional support
from the John D. and Catherine T. MacArthur Foundation and the Peter and Patricia Gruber Foundation.
DKE was supported by the National Aeronautics and Space Administration under Award No. NAS7-03001 and the
California Institute of Technology. Gwen Rudie and Olivera Rakic each provided  very helpful
comments on an earlier draft of the paper; we also thank Kurt Adelberger for his early involvement
in the work which made the new results possible. We are grateful 
to the staff of the W.M. Keck Observatory who keep the instruments
and telescopes running effectively. A careful reading and constructive report by the referee is very
much appreciated. Finally,
we wish to extend thanks to those of Hawaiian ancestry on whose sacred mountain we are privileged
to be guests.

\bibliographystyle{apj}
\bibliography{refs}

\end{document}